\documentclass[prd,aps,eqsecnum,floatfix,nofootinbib,preprint,tightenlines]{revtex4}

\usepackage{latexsym}
\usepackage{graphicx}
\usepackage[dvipsnames]{xcolor}

\def\bibi{\bibitem}

\let\inodot=\i

\def\a{\alpha}
\def\b{\beta}
\def\c{\chi}
\def\d{\delta}
\def\g{\gamma}

\def\i{\iota}

\def\m{\mu}
\def\n{\nu}
\def\o{\omega}
\def\p{\pi}                     
\def\r{\rho}                    
\def\s{\sigma}                  
\def\t{\tau}

\def\D{\Delta}

\def\P{\Pi}

\def\cbo{{\,\raise-.15ex\Sc [\,}}                       
\def\ltap{\raisebox{-.4ex}{\rlap{$\sim$}} \raisebox{.4ex}{$<$}}   

\def\ddt#1{{\buildrel {\hbox{\LARGE .\kern-2pt.}} \over {#1}}}

\def\ie{\mbox{\it i.e.}}

\def\etc{\mbox{\it etc.}}

\def\floatcaption#1#2{ \caption{ #2 \ [#1] \label{#1}} }
\def\floatcaption#1#2{ \caption{#2 \label{#1}} }

\long\def\symbolfootnote[#1]#2{\begingroup%
\def\thefootnote{\fnsymbol{footnote}}\footnote[#1]{#2}\endgroup}

\long \def \blockcomment #1\endcomment{}

\def\seef{{\it cf.}}

\def\bq{{\overline{q}}}

\begin{document}

\thispagestyle{empty}

\begin{center}
\vspace*{5mm}
\begin{boldmath}
{\large\bf The strong coupling from hadronic $\t$ decays: \\
a critical appraisal}
\end{boldmath}
\\[10mm]
Diogo Boito,$^a$
Maarten Golterman,$^b$
Kim Maltman,$^{c,d}$ Santiago Peris$^e$
\\[8mm]
{\small\it
$^a$Instituto de F{\'\inodot}sica de S{\~a}o Carlos, Universidade de S{\~a}o Paulo\\
CP 369, 13570-970, S{\~a}o Carlos, SP, Brazil
\\[5mm]
$^b$Department of Physics and Astronomy,
San Francisco State University\\ San Francisco, CA 94132, USA
\\[5mm]
$^c$Department of Mathematics and Statistics,
York University\\  Toronto, ON Canada M3J~1P3
\\[5mm]
$^d$CSSM, University of Adelaide, Adelaide, SA~5005 Australia
\\[5mm]
$^e$Department of Physics and IFAE-BIST, Universitat Aut\`onoma de Barcelona\\
E-08193 Bellaterra, Barcelona, Spain}
\\[10mm]
\end{center}

\begin{quotation}
Several different analysis methods have been developed to determine
the strong coupling via finite-energy sum-rule analyses of hadronic 
$\t$ decay data. While most methods agree on the existence of 
the well-known ambiguity in the choice of a resummation scheme 
due to the slow convergence of QCD perturbation theory 
at the $\t$ mass, there is an ongoing controversy over how
to deal properly with non-perturbative effects. These
 are small, but not negligible, and include quark-hadron 
``duality violations'' (\ie, resonance effects) which are not 
described by the operator product expansion (OPE). In one approach, 
an attempt is made to suppress duality violations enough that
they might become negligible. The number of OPE parameters to be fit, 
however, then exceeds the number of available sum rules, necessitating 
an uncontrolled
OPE truncation, in which a number of higher-dimension
OPE contributions in general present in QCD are set to zero by hand.
In the second approach, truncation of the OPE is avoided by 
construction, and duality violations are taken into account explicitly, 
using a physically motivated model. In this article, we provide 
a critical appraisal of a recent analysis employing the first 
approach and demonstrate that it fails to properly account for 
non-perturbative effects, making the resulting determination of the 
strong coupling unreliable. The second approach, in contrast, passes all 
self-consistency tests, and provides a competitive 
determination of the strong coupling from $\t$ decays.
\end{quotation}

\newpage
\section{\label{introduction} Introduction}
A precise determination of the strong coupling $\a_s$ is important, 
both because it is one of the fundamental parameters of the Standard Model, 
and because it is an important input to precision studies of potential 
discrepancies between experiment and theory, relevant to searches
for beyond-the-Standard-Model physics. Moreover, determinations over a 
wide range of energies provide an important test of the running of the 
coupling as predicted by QCD.

Experimental data for hadronic $\t$ decays provide an opportunity for a 
determination at quite low energy scales, of order the $\t$ mass. Because 
of the long running from the $\t$ mass to the $Z$ mass, even a modestly 
accurate determination translates into a high-precision value at the $Z$ 
mass, and, as such, provides a stringent test of QCD. However, $\a_s$ is, 
of course, defined in perturbation theory, and it is thus imperative to 
have a quantitative understanding of non-perturbative effects that 
may ``contaminate'' determinations at lower scales such as that at 
the $\t$ mass, where resonance effects are clearly visible in 
the QCD spectral functions extracted from differential $\t$ decay
distributions. Such resonance effects, which are described neither by
perturbation theory nor by the operator product expansion (OPE), 
are referred to generically in the literature as ``violating quark-hadron duality.''\footnote{The term ``quark-hadron duality'' is shorthand for
the qualitative expectation that QCD spectral functions, at least in some 
average sense, can be equally well understood in terms of quarks and 
gluons (the perturbative picture) as in terms of a tower of resonances 
(the non-perturbative picture).} A quantitative study of the impact 
of duality violations (DVs) is unavoidable if one aims to fully understand 
the possible systematics affecting the extraction of $\a_s(m_\t^2)$ from 
$\t$ decays.

Two basic strategies have been developed to extract $\a_s(m_\t^2)$
from hadronic $\t$ decay data. Both are based on the use of finite-energy sum rules (FESRs), in 
which weighted integrals of the vector ($V$) and axial-vector ($A$) 
hadronic spectral functions (or their sum) are related to the integral 
on a circle around the origin in the complex plane over a theoretical
representation of the $V$ or $A$ current two-point function. Choosing 
the radius $s_0$ of this circle to be large makes it possible to use 
perturbation theory, augmented by the OPE and possibly also
with a model for residual DVs, for the theoretical representation, 
allowing for the extraction of $\a_s(s_0)$. Of course, in the application
of FESRs to $\t$ decays, the maximum radius is $s_0=m_\t^2$.

In the first strategy, the weights in the spectral integrals are chosen 
with the hope of suppressing DVs enough to justify omitting them 
from the analysis, and, in this spirit, $s_0$ is always chosen equal to 
its maximum kinematically allowed value, $m_\t^2$. As we will see, 
in order to keep the number of the resulting sum rules greater
than the number of parameters to be fit, this choice forces a truncation
of the OPE at a dimension lower than a complete QCD analysis
would generally require. We will refer to this strategy as the 
``truncated-OPE-model'' or ``truncated-OPE'' strategy. The most recent 
implementations of this strategy can be found in Refs.~\cite{ALEPH13,Pich}.

In the second strategy, weights are chosen such that only low 
orders in the OPE need to be included. It turns out that this is 
incompatible with the desired complete suppression of DVs, 
and an explicit model of how they affect the spectral functions 
needs to be introduced in order to carry out the analysis. The value of 
$s_0$ is varied between approximately $1.5$~GeV$^2$ and $m_\t^2$. 
This approach has been followed in Refs.~\cite{alphas1,alphas2,alphas14}, 
and we will refer to it as the ``DV-model strategy.''

Both strategies are based on assumptions, and these assumptions have to be
tested. This is not an academic issue, because the values of $\a_s(m_\t^2)$
obtained by applying the two different strategies to the same 
data set differ significantly, the DV-model result being lower by 
about $8\%$. This difference is a factor of two larger than the 
$4\%$ (or less) errors produced using the individual strategies.

The goal of  this article is to present a critical analysis of the 
truncated-OPE-model strategy, starting from the extensive analysis 
recently presented in Ref.~\cite{Pich}. In Ref.~\cite{Pich} a large number of 
tests of this strategy were carried out, leading to the claim that
the strategy is robust, even if there is no good {\it a priori}
physical motivation for the truncation of the OPE employed.
Here we will demonstrate that despite these tests, this strategy does 
not, in fact, hold up, and that consequently the final result for 
$\a_s(m_\t^2)$ obtained in Refs.~\cite{ALEPH13,Pich} is unreliable. In 
particular, while the tests carried out in Ref.~\cite{Pich} are certainly 
necessary, they are not sufficient to be confident that the systematic 
errors quoted in Ref.~\cite{Pich}, following from the use of this strategy, 
are under control. In what follows, we will show explicitly that 
they are not.

Of course, the DV-model strategy requires similar scrutiny, and
numerous self-consistency tests have already been carried out in 
Refs.~\cite{alphas1,alphas2,alphas14}. The details of these tests will 
not be repeated here, but may be found in those references. As we will 
argue below, the deficiencies of the truncated-OPE-model strategy in 
fact naturally lead one to adopt the DV-model strategy, a point already 
made in some detail in Ref.~\cite{alphas1}. 
We will also show that the criticism of 
the DV-model strategy in Ref.~\cite{Pich} is misleading, and in
fact in no way invalidates the DV-model strategy approach.

This article is organized as follows. In the next section, we collect 
elements of the theory of FESRs needed for our purposes. Then, in 
Sec.~\ref{PRresults}, we begin by reproducing the results of Ref.~\cite{Pich}, 
and end with a discussion of hints of instabilities already visible in 
these results. In Sec.~\ref{fake data}, we carry out a numerical experiment 
using fake data which are compatible with the experimental spectral functions, and which have been
generated from a model of the $V$ and $A$ spectral functions with fixed 
input $\a_s(m_\t^2)$ and, by construction, non-negligible DVs. We show 
that fits extracting $\a_s(m_\t^2)$ from these data employing the 
truncated-OPE strategy fail to reproduce the exactly known model 
value of $\alpha_s(m_\tau^2)$ by an amount comparable to
the difference found when the two strategies are applied to the real data.
In Sec.~\ref{DVdiscussion} we refute the critique of the
DV-model strategy contained in Ref.~\cite{Pich}. Section~\ref{conclusion}
contains our conclusions.

\section{\label{theory} Theory}
The sum-rule analysis underlying both strategies starts from the 
current-current 
two-point functions
\begin{eqnarray}
\label{correl}
\P_{\m\n}(q)&=&i\int d^4x\,e^{iqx}\langle 0|T\left\{J_\m(x)J^\dagger_\n(0)\right\}|0\rangle\\
&=&\left(q_\m q_\n-q^2 g_{\m\n}\right)\P^{(1)}(q^2)+q_\m q_\n\P^{(0)}(q^2)\nonumber\\
&=&\left(q_\m q_\n-q^2 g_{\m\n}\right)\P^{(1+0)}(q^2)+q^2 g_{\m\n}\P^{(0)}(q^2)\ ,\nonumber
\end{eqnarray}
where $J_\m$ is the non-strange $V_\m=\overline{u}\g_\m d$
or $A_\m=\overline{u}\g_\m\g_5 d$ current, and
the superscripts $(0)$ and $(1)$ label spin. The combinations $\P^{(1+0)}(q^2)$ and $q^2\P^{(0)}(q^2)$ are free of kinematic singularities.
Defining $s=q^2=\, -Q^2$ and the spectral function
\begin{equation}
\label{spectral}
\r^{(1+0)}(s)=\frac{1}{\p}\;\mbox{Im}\,\P^{(1+0)}(s)\ ,
\end{equation}
Cauchy's theorem applied to the contour in Fig.~\ref{cauchy-fig}
and the analytical properties of $\P^{(1+0)}(s)$
 imply the FESR
\begin{eqnarray}
\label{cauchy}
\frac{1}{s_0}\int_0^{s_0}ds\,w(s/s_0)\,\r^{(1+0)}_{V/A}(s)
&=&-\frac{1}{2\p i\, s_0}\oint_{|s|=s_0}
ds\,w(s/s_0)\,\P^{(1+0)}_{V/A}(s)\ .
\end{eqnarray}
This FESR is valid for any $s_0>0$ and any weight $w(s)$ analytic 
inside and on the
contour \cite{shankar,Braaten88,BNP}.   
It holds for the $V$ and $A$ cases
separately, and as a consequence also for $V+A$.

\begin{figure}
\vspace*{4ex}
\begin{center}
\includegraphics*[width=6cm]{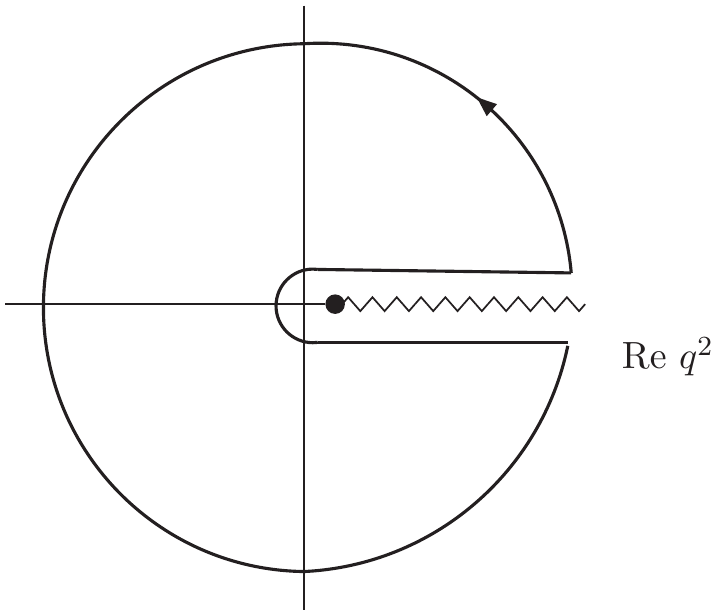}
\end{center}
\begin{quotation}
\floatcaption{cauchy-fig}%
{{\it Analytic structure of $\P^{(1+0)}(q^2)$ in the complex $s=q^2$ plane.
There is a cut on the positive real axis starting at $s=q^2=4m_\p^2$ 
(a pole at $s=q^2=m_\p^2$ and a cut starting at $s=9m_\p^2$) for the $V$ 
($A$) case. The solid curve shows the contour used in Eq.~(\ref{cauchy}).}}
\end{quotation}
\vspace*{-4ex}
\end{figure}

Experimental results for the  flavor $ud$ $V$ and $A$ spectral functions have 
been made available by ALEPH and OPAL in Refs.~\cite{ALEPH13,ALEPH,OPAL}.
Apart from the pion-pole contribution,  
$\rho_{V/A;ud}^{(0)}(s) = O[(m_d\mp m_u)^2]$ is chirally
suppressed, and
the continuum part of $\rho_{V/A}^{(0)}(s)$ is thus numerically negligible.

For large $|s|=s_0$, and far enough away from the positive real axis,
$\P^{(1+0)}(s)$ can be approximated by the OPE
\begin{equation}
\label{OPE}
\P^{(1+0)}_{\rm OPE}(s)=\sum_{k=0}^\infty \frac{C_{2k}(s)}{(-s)^{k}}\ .
\end{equation}
We will omit the labels $V$, $A$ or $V+A$ on the OPE coefficients, because
we will only encounter the case $V+A$ in this article.
The $C_{2k}$ are logarithmically dependent on $s$ through
perturbative corrections.
The term with $k=0$ corresponds
to the purely perturbative, mass-independent contributions, which have been
calculated to order $\a_s^4$ in Ref.~\cite{PT}, and are the
same for the $V$ and $A$ channels. Values quoted
for $\a_s(m_\t^2)$ are in the $\overline{\rm MS}$ scheme. We will
consider both FOPT and CIPT \cite{CIPT} resummation schemes
in evaluating the truncated perturbative series (see for 
instance Refs.~\cite{MJ,BJ} for a discussion of these two resummation schemes).
The $C_{2k}$ with $k\ge 1$ are different for the $V$ and $A$ channels, 
and, for $k>1$, contain non-perturbative $D=2k$ condensate contributions. 
As in Refs.~\cite{ALEPH13,Pich,alphas1,alphas2,alphas14}, we will neglect 
purely perturbative quark-mass contributions to $C_2$ and $C_4$, as they 
are numerically very small for the non-strange FERSs under consideration. 
We will also neglect the $s$-dependence of the coefficients $C_{2k}$ for 
$k>1$, because they are $\a_s$ suppressed. With this choice, we have that
\begin{equation}
\label{degreecond}
\frac{1}{2\p is_0}\oint_{|s|=s_0}\,ds\,\left(\frac{s}{s_0}\right)^n
\,\frac{C_{2k}}{(-s)^k} =(-1)^{n+1}\frac{C_{2(n+1)}}{s_0^{n+1}}\,\d_{k,n+1}\ ,
\end{equation}
implying that an $n$-th degree monomial in the weight $w(s/s_0)$ selects the
$D=2k=2(n+1)$ term in the OPE.

All fit results for the $1/s^4$ term in the OPE will be given in terms of 
$C_4$, while Ref.~\cite{Pich} chose to use the gluon condensate, 
$\langle\frac{\a_s}{\p}GG\rangle$, instead. For the $V+A$ case, these two 
parameters are related by
\begin{equation}
\label{c4}
C_4=\frac{1}{6}\left(1-\frac{11}{18}\frac{\a_s}{\p}\right)
\left\langle\frac{\a_s}{\p}GG\right\rangle
+2\left(1-\frac{23}{27}\frac{\a_s}{\p}\right)\langle(m_u+m_d)\bq q\rangle\ .
\end{equation}
If we employ, as in Ref.~\cite{Pich}, the value 
$\langle(m_u+m_d)\bq q\rangle\approx -m_\p^2 f_\p^2\approx -1.6\times
10^{-4}$~GeV$^4$ with $\a_s\approx 0.325$, Eq.~(\ref{c4}) translates into
\begin{equation}
\label{c4num}
C_4=0.156\,\left\langle\frac{\a_s}{\p}GG\right\rangle-0.000292\ \mbox{GeV}^4\ .
\end{equation}

Perturbation theory, and, more generally, the OPE, breaks down near the 
positive real axis \cite{PQW}. In order to account for this, we replace 
the right-hand side of Eq.~(\ref{cauchy}) by
\begin{equation}
\label{split}
-\frac{1}{2\p is_0}\oint_{|s|=s_0}ds\,w(s/s_0)\,
\left(\P^{(1+0)}_{\rm OPE}(s)+\D(s)\right)\ ,
\end{equation}
with
\begin{equation}
\label{DVdef}
\D(s)\equiv\P^{(1+0)}(s)-\P^{(1+0)}_{\rm OPE}(s)\ ,
\end{equation}
where the difference $\D(s)$ represents, by definition, the
quark-hadron duality violating contribution to $\Pi^{(1+0)}(s)$.
As shown in Ref.~\cite{CGP}, the integral over $w(s/s_0)\D(s)$ in
Eq.~(\ref{split}) can be rewritten such that the FESR takes the form
\begin{eqnarray}
\label{sumrule}
&&\hspace{-1cm}\frac{1}{s_0}\int_0^{s_0}ds\,w(s/s_0)\,\r^{(1+0)}_{V/A}(s) \\
&&=
-\frac{1}{2\p is_0}\oint_{|s|=s_0}
ds\,w(s/s_0)\,\P^{(1+0)}_{{\rm OPE},V/A}(s)-\frac{1}{s_0}\,
\int_{s_0}^\infty ds\,w(s/s_0)\,\frac{1}{\p}\,\mbox{Im}\,
\D_{V/A}(s)\ ,\nonumber
\end{eqnarray}
if $\D(s)$ is assumed to decay fast enough as $s\to\infty$. The imaginary
parts $\frac{1}{\p}\,\mbox{Im}\,\D_{V/A}(s)$ can thus be interpreted as
the duality-violating parts, $\rho_{V/A}^{\rm DV}(s)$, of the $V/A$ spectral functions.

We need to resort to a model in order to account
for DVs, because the functional form of $\D(s)$ is not known, even for large
$s$. As in Refs.~\cite{CGP,CGPmodel,CGP05},\footnote{See
also Refs.~\cite{russians,MJ11}.} our model is based on large-$N_c$ and Regge
considerations, parametrizing $\rho_{V/A}^{\rm DV}(s)$
as
\begin{equation}
\label{ansatz}
\r_{V/A}^{\rm DV}(s)=
e^{-\d_{V/A}-\g_{V/A}s}\sin{(\a_{V/A}+\b_{V/A}s)}\ ,\qquad s\ge s_{\rm min}\ .
\end{equation}
This introduces four new parameters in each channel, 
in addition to $\a_s$ and the $D\ge 4$ OPE coefficients.\footnote{An {\it ansatz} of the 
form~(\ref{ansatz}) was used by the authors of Ref.~\cite{Pich} to model DVs in the
$V-A$ spectral function \cite{LEC}. The difference 
$\r_V^{\rm DV}(s)-\r_A^{\rm DV}(s)$ was used instead in Ref.~\cite{BetalLEC}.}
Our {\it ansatz}~(\ref{ansatz}) is assumed to hold only for 
$s\ge s_{\rm min}$, with $s_{\rm min}$ to be determined from fits to the data.
We emphasize that, since DVs represent resonance effects, the DV 
parameters will be different in the $V$ and $A$ channels, reflecting 
the different resonance structure in these two channels.

One way the general structure of the {\it ansatz} proposed in Eq.~(\ref{ansatz}) 
can be understood is as follows. The OPE itself diverges as an expansion in 
$1/s$, because it has zero radius of convergence around $1/s=0$. It is 
thus itself, like the perturbative series in the $D=0$ term \cite{MB}, 
at best an asymptotic expansion, with coefficients $C_{2k}$ that must 
eventually grow rapidly with $k$.  This then leads to the expectation
that an exponential correction suppressed in terms of the inverse of 
this expansion parameter, ``takes over'' where the OPE starts to diverge.
The form given in Eq.~(\ref{ansatz}) is consistent with this expectation. 
Moreover, following this line of reasoning, it would be natural to 
expect that the prefactor of the form~(\ref{ansatz}) is itself an 
expansion in powers of $1/s$. In {\it ansatz}~(\ref{ansatz}) only the
leading (constant) term was kept in this prefactor expansion.

\section{\label{PRresults} The truncated-OPE-model strategy}
In this section, we will first summarize the truncated-OPE strategy 
of Ref.~\cite{Pich}. We will then, after reproducing the results 
of Ref.~\cite{Pich} in Sec.~\ref{reproduction}, start a critical discussion of both 
these results and the underlying strategy in Sec.~\ref{discussion3}.

The orginal version \cite{DP1992} of the truncated-OPE strategy
employed five different FESRs, corresponding to five different polynomial
choices for the weight function $w(x)$ in the FESR~(\ref{cauchy}).
We will denote these weights as $w_{k\ell}$, with $(k\ell)\in\{(00),(10),(11),(12),(13)\}$, and
\begin{equation}
\label{ALEPH}
w_{k\ell}(x)=(1-x)^{k+2}x^\ell(1+2x)\ .
\end{equation}
These weights have a double or triple zero at $s=s_0$ (\ie, $x=1$), and the 
hope was that this would be sufficient to suppress DVs enough 
that they could be neglected.\footnote{We will refer to a weight 
with an $n$-fold zero at $s=s_0$ as $n$-fold pinched \cite{KM98,DS99}.}
In other words, in all the fits of Ref.~\cite{Pich} all weighted 
integrals of the functions $\r^{\rm DV}_{V/A}(s)$ of Eq.~(\ref{ansatz}) 
are set equal to zero. In practice, this is equivalent to choosing 
a model in which $\r^{\rm DV}_{V/A}(s)=0$,
\ie, $\d_{V/A} = \infty$, from the outset, regardless of the oscillations
clearly visible in Fig.~\ref{V+A-1}. The value of $s_0$ was chosen equal to $m_\t^2$, again with the hope 
that this would maximize the suppression of non-perturbative effects 
represented by the $D=2k>0$ terms in the OPE, Eq.~(\ref{OPE}), and $\D(s)$, 
Eq.~(\ref{DVdef}). {\it Prima facie}, this is not an implausible 
assumption, as perturbation theory should provide an approximation to
the right-hand side of Eq.~(\ref{cauchy}) that becomes more accurate as
$s_0$ increases.

These choices imply that one has five data points, and one thus needs to
limit the number of parameters in the fit to four (or less). This leads to
the necessity of truncating the OPE. Since $C_2$ is negligibly small for
the non-strange channels considered here, the choice was made to
take $\a_s(m_\t^2)$, $C_4$, $C_6$ and $C_8$ as free parameters in the
fit. However, because of Eq.~(\ref{degreecond}), this amounts to the additional
assumption that $C_{10}=C_{12}=C_{14}=C_{16}=0$, since the set
Eq.~(\ref{ALEPH}) contains polynomials with degree up to seven. While such an
assumption is necessary to implement the chosen fit strategy, 
it has no basis in QCD. 

In addition to the set of weights in Eq.~(\ref{ALEPH}), Ref.~\cite{Pich} considered 
several other sets of polynomials in $x$, in order to test this strategy. 

In one set, which we refer to as the ``reduced set'', denoted
$w_{k\ell}^{\rm red}(x)$, the factor $1+2x$ was removed from the $w_{k\ell}$.
The form of the weights $w_{k\ell}^{\rm red}(x)$ is then
\begin{equation}
\label{reduced}
w^{\rm red}_{k\ell}(x)=(1-x)^{k+2}x^\ell\ .
\end{equation}
Again, the pair $(kl)$ was chosen in the set
$\{(00),(10),(11),(12),(13)\}$. The motivation for this 
choice is that it ``reduces'' the number of assumptions associated with the 
chosen OPE truncation; one needs only assume $C_{10}=C_{12}=C_{14}=0$, 
since $C_{16}$ is not probed by this modified set. Other sets 
of weights can be chosen from among the ``optimal'' weights 
$w_{m,n}^{\rm opt}(x)$, where
\begin{equation}
\label{optimal}
w^{\rm opt}_{m,n}=(1-x)^{1+m}\left(\frac{d~}{dx}\right)^m
\sum_{k=0}^n x^{m+k}\ ,\qquad m=0, 1\ ,\qquad n=1,\dots,5\ .
\end{equation}
For $m=0$ each of these weights selects only one $D>0$ term in the OPE,
and for $m=1$ each of these weights selects only two $D>0$ terms.
The $m=0$ ``optimal'' weights are singly pinched and the $m=1$ ``optimal''
weights are doubly pinched. The most important of these weights are
those with $m=1$, because they are doubly pinched, and, for $n\ge 1$
the two OPE terms probed by these weights have $D\ge 6$, thus
avoiding a contribution from the nominally dominant $D=4$ term.
The $m=0$ weights probe only one term in the OPE, but are expected to be less
effective in suppressing DVs because they are only singly 
pinched.\footnote{One should bear in mind, however, that a higher degree 
of pinching does not always guarantee a stronger suppression of DVs, 
when one is considering only a single value of $s_0$ \cite{mainz2}.}
For the set with $m=1$ and $1\le n\le 5$, the truncation
assumption amounts to setting $C_{12}=C_{14}=C_{16}=0$.

Yet another set of weights considered was the four-weight set, $w_n(x)$, 
$n=0,\dots ,3$, with
\begin{equation}
\label{lesspinched}
w_n(x)=(1-x)^n\ .
\end{equation}
For this set, the parameters $\a_s(m_\t^2)$, $C_4$ and $C_6$ were fit, 
while the OPE was truncated by setting $C_8=0$ in order to have
one degree of freedom in the fit, even though $C_8$ is probed by $w_3$.
This set is a little different in nature, because $w_0$ is unpinched, 
and $w_1$ is only singly pinched. Therefore, the effects from DVs are 
potentially more severe.

The truncation assumption affects the higher-degree weights more.
For example, $w_{00}=w^{\rm opt}_{1,1}$ is not affected, 
because it only probes OPE terms with $D\le 8$, $w_{10}$ and 
$w^{\rm opt}_{1,2}$ only probe in addition the $D=10$ term, 
\etc\ This could lead one to hope that, despite the fact that 
the assumed truncation has no ground in QCD, the determination of
$\a_s$ might be less severely affected. This could, for
example, happen if the spectral moments involving lower-degree
weights are relatively more important in fixing $D=0$ contributions 
than are those involving higher-degree weights. Any such speculation
should, of course, be explicitly tested. While Ref.~\cite{Pich} carried 
out a number of such tests (which we will also consider below), 
we will nevertheless see that the truncation assumption employed
in the truncated-OPE strategy has a significant impact on the value of 
$\a_s$ obtained from this collection of fits. In other words, though
the many tests in Ref.~\cite{Pich} can be considered as necessary, 
they turn out not to be sufficient.
\subsection{\label{reproduction} Reproduction of the fits of Ref.~\cite{Pich}}
Before we investigate the validity of the truncated-OPE strategy, we first
reproduce the fits of Ref.~\cite{Pich} based on this strategy. We will only 
consider fits to moments computed from the sum of the $V$ and $A$ non-strange 
spectral functions, as Ref.~\cite{Pich} advocates that this is the most reliable 
choice. We primarily consider fits using the four sets of weights specified 
above, and our version of the results is reported in Tables~\ref{tab1}, 
\ref{tab2}, \ref{tab3} and \ref{tab4} below.

These are all good fits, and the results agree with those found 
in Ref.~\cite{Pich}, within our statistical errors. Since our goal is not 
to obtain final results for $\a_s$ from these fits, we do not repeat 
the estimates of systematic errors carried out in Ref.~\cite{Pich}. We note 
that the central values are slightly different. This is most likely due 
to a slightly different treatment of the data, including a small rescaling 
performed in Ref.~\cite{alphas14}.\footnote{This rescaling was required
in order to restore the correct total non-strange normalization, fixed by
the electron, muon and total strange branching fractions, after the 
larger-error experimental $\tau$ decay $\pi$ pole strength was replaced 
by the more precise value implied by $\pi_{\mu2}$ and the Standard Model.
For details of our treatment of the data, see Ref.~\cite{alphas14}.}
Using Eq.~(\ref{c4num}), it is straightforward to verify that our values 
for $C_4$ are consistent with the values for $\langle\frac{\a_s}{\p}GG\rangle$
 given in Ref.~\cite{Pich}. We verified the results found in Table~5 of 
Ref.~\cite{Pich} as well, with similar accuracy. We do not show these here,
since they are less central to the final value for $\a_s(m_\t^2)$ 
quoted in Ref.~\cite{Pich}.
\begin{table}[h!]
\hspace{0cm}\begin{tabular}{|c|c|c|c|c|c|}
\hline
& $\a_s(m_\t^2)$ & $C_{4}$ (GeV$^4$) & $C_{6}$ (GeV$^6$)& $C_{8}$ (GeV$^8$)& $\c^2/$dof \\
\hline
FOPT & 0.316(3) & -0.0006(3) & 0.0012(3) & -0.0008(3) & 1.38/1 \\
\hline
CIPT & 0.336(4) & -0.0026(4)  & 0.0009(3) & -0.0010(4) & 0.89/1 \\
\hline
\end{tabular}
\floatcaption{tab1}{{\it Reproduction of the $V+A$ fits of Table 1 of 
Ref.~\cite{Pich}, based on the weights of Eq.~(\ref{ALEPH}).
By assumption, $C_{10}=C_{12}=C_{14}=C_{16}=0$.
Errors are statistical only.}}
\end{table}%
\begin{table}[h!]
\hspace{0cm}\begin{tabular}{|c|c|c|c|c|c|}
\hline
& $\a_s(m_\t^2)$ & $C_{4}$ (GeV$^4$) & $C_{6}$ (GeV$^6$)& $C_{8}$ (GeV$^8$)& $\c^2/$dof \\
\hline
FOPT & 0.316(2) & -0.0005(1) & 0.0011(1) & -0.0005(1) & 1.57/1 \\
\hline
CIPT & 0.336(4) & -0.0025(3)  & 0.0008(2) & -0.0008(2) & 0.98/1 \\
\hline
\end{tabular}
\floatcaption{tab2}{{\it Reproduction of the $V+A$ fits of Table 3 of 
Ref.~\cite{Pich}, based on the reduced weights~(\ref{reduced}). By assumption, 
$C_{10}=C_{12}=C_{14}=0$.
Errors are statistical only.}}
\end{table}%
\begin{table}[h!]
\hspace{0cm}\begin{tabular}{|c|c|c|c|c|c|}
\hline
& $\a_s(m_\t^2)$  & $C_{6}$ (GeV$^6$)& $C_{8}$ (GeV$^8$) & $C_{10}$ (GeV$^{10}$)& $\c^2/$dof \\
\hline
FOPT & 0.317(3)  & 0.0014(4) & -0.0010(5) & 0.0004(3) & 1.26/1 \\
\hline
CIPT & 0.336(4)   & 0.0010(4) & -0.0011(5) & 0.0003(3) &  0.83/1 \\
\hline
\end{tabular}
\floatcaption{tab3}{{\it Reproduction of the $V+A$ fits of Table 7 of 
Ref.~\cite{Pich}, based on the ``optimal'' weights~(\ref{optimal}) with 
$m=1$ and $n=1,\dots\,5$. By assumption, $C_{12}=C_{14}=C_{16}=0$.
Errors are statistical only.}}
\end{table}%
\begin{table}[h!]
\hspace{0cm}\begin{tabular}{|c|c|c|c|c|c|}
\hline
& $\a_s(m_\t^2)$  & $C_{4}$ (GeV$^4$)& $C_{6}$ (GeV$^6$) & $\c^2/$dof \\
\hline
FOPT & 0.320(8) & -0.001(1) & 0.002(2)  & 1.25/1 \\
\hline
CIPT & 0.339(11) & -0.003(3)  & 0.001(2)  & 1.15/1 \\
\hline
\end{tabular}
\floatcaption{tab4}{{\it Reproduction of the $V+A$ fits of Table 6 of 
Ref.~\cite{Pich}, based on the weights of Eq.~(\ref{lesspinched}) with 
$n=0,\dots\,3$. By assumption, $C_{8}=0$.
Errors are statistical only.}}
\end{table}%
\subsection{\label{discussion3} Critique}
We will now turn to a discussion of observations on the basic assumptions
underlying the truncated-OPE-model strategy, based on the data. First, we 
consider the OPE truncation itself, then investigate the other 
key assumption that, at the $\t$ scale, double (or triple) 
pinching produces a suppression of DV contributions strong enough to
allow them to be ignored.
\subsubsection{\label{truncation} Truncation of the OPE}

To obtain the results reported in Tables~\ref{tab1} to \ref{tab4} above, two 
major assumptions have been made. The first is that setting to 
zero by hand higher-dimension OPE contributions in principle present in
the analysis (unavoidable if one wishes to have at least one degree of 
freedom in the fit) has no significant impact on the resulting  
$\alpha_s(m_\tau^2)$.
This assumption was tested
in Ref.~\cite{Pich} by relaxing this constraint on the coefficient $C_D$ with 
$D$ equal to the lowest dimension of the OPE term neglected in the fits 
described in Sec.~\ref{reproduction}. For the fits of Table~\ref{tab1} and 
\ref{tab2} this means that now also $C_{10}$ is left as a free
parameter, while for Table~\ref{tab3} the corresponding new free
parameter is $C_{12}$. Of course, now we have no degrees of freedom left, 
the minimal value of $\c^2$ is zero, and these tests are not proper fits. 
Nonetheless, errors on the free parameters can still be found through 
linear error propagation, and these results can thus be compared with the 
results reported in Sec.~\ref{reproduction}.

Here, let us reproduce the first example of these tests. We again carry out 
a ``fit'' to the spectral-function integrals with weights~(\ref{ALEPH}), 
but now use these to determine the parameters $\a_s$, $C_{4,6,8}$
{\it and} $C_{10}$. Our results are reported in Table~\ref{tab5}.
\begin{table}[h!]
\hspace{0cm}\begin{tabular}{|c|c|c|c|c|c|}
\hline
& $\a_s(m_\t^2)$ & $C_{4}$ (GeV$^4$) & $C_{6}$ (GeV$^6$)& $C_{8}$ (GeV$^8$)& $C_{10}$ (GeV$^{10}$) \\
\hline
FOPT & 0.329(12) & -0.0014(8) & 0.005(4) & -0.004(3) & 0.010(8) \\
\hline
CIPT & 0.350(15) & -0.0036(12)  & 0.004(3) & -0.004(3) & 0.007(8) \\
\hline
\end{tabular}
\floatcaption{tab5}{{\it Reproduction of the $V+A$ ``fits'' of Table 2 of 
Ref.~\cite{Pich}, based on the weights of Eq.~(\ref{ALEPH}), to be compared with 
Table~\ref{tab1}. By assumption, $C_{12}=C_{14}=C_{16}=0$, while $C_{10}$ 
is now left as a free parameter.
Errors are obtained through linear error propagation.}}
\end{table}%

These results are in agreement with Table~2 of Ref.~\cite{Pich}, within errors.
They are also in agreement within errors with Table~\ref{tab1}. 
Ref.~\cite{Pich} takes this as a sign of stability of the fits of 
Table~\ref{tab1}, and thus as a validation of the truncation of the 
OPE beyond the $D=8$ term. However, while the errors on the coefficients 
$C_6$ and $C_8$ are large, so that there is no inconsistency between the 
values of these parameters obtained in Tables~\ref{tab1} and \ref{tab5}, 
one notes that their central values in Table~\ref{tab5} are roughly five 
times as large as those of Table~\ref{tab1}.
We also note that the central values of $\a_s(m_\t^2)$ in Table~\ref{tab5}
are larger than in Tables~\ref{tab1} to \ref{tab4}.
Similar observations hold for similar ``fits'' with no degrees of freedom 
with reduced and ``optimal'' weights,
\seef\ Tables~4 and 8 in Ref.~\cite{Pich}.\footnote{In some cases, the factor is
closer to ten than to five. We have reproduced the $V+A$ results of these
tables.} This suggests that the OPE coefficients may ``want'' to be larger, 
but that just adding one more term in the fit, while still truncating the 
remaining terms, does not allow the OPE the ``room'' to do this. As we will 
see below, reasonable values for the OPE condensates exist which are 
compatible with the data, but which lead to significantly lower 
values of $\alpha_s$. The tests carried out in Ref.~\cite{Pich}, and given in 
their Tables~2, 4 and 8, are thus, in fact, inconclusive.

Interestingly, no such test was carried out for the fits of Table~6 in 
Ref.~\cite{Pich}, which we reproduce here in Table~\ref{tab4}. Such a test 
can, of course, be performed by leaving $C_8$ as a free parameter. 
We carried out this test, and find the values in Table~\ref{tab6} below. We 
note that there is a very dramatic shift in the central value of 
$\a_s(m_\t^2)$, of about 24\%, while also the errors increase dramatically. 
Taken all together, we conclude that tests based on ``fits'' with zero 
degrees of freedom add no information, and are certainly not a
demonstration of stability of the truncated-OPE strategy.

\begin{table}[h!]
\hspace{0cm}\begin{tabular}{|c|c|c|c|c|}
\hline
& $\a_s(m_\t^2)$ & $C_{4}$ (GeV$^4$) & $C_{6}$ (GeV$^6$)& $C_{8}$ (GeV$^8$)\\
\hline
FOPT & 0.39(6) & -0.02(3) & 0.07(8)  & -0.2(2) \\
\hline
CIPT & 0.43(10) & -0.03(3)  & 0.06(6)  & -0.2(2)  \\
\hline
\end{tabular}
\floatcaption{tab6}{{\it Test of the $V+A$ ``fits'' of Table 6 of 
Ref.~\cite{Pich}, based on the weights of Eq.~(\ref{lesspinched}), to be compared 
with Table~\ref{tab4}. $C_{8}$ is now left as a free parameter.
Errors are obtained through linear error propagation.
}}
\end{table}%

\bigskip
We now will consider an exercise which shows that the whole collection of 
fits carried out in Ref.~\cite{Pich} admits a very different solution. This
will serve to demonstrate that the argument of
Ref.~\cite{Pich}, that the consistency of the results obtained 
from all the tests performed there establishes the robustness of the 
determination of $\a_s$, is false.

Let us return to the fits of Tables~\ref{tab1} to \ref{tab4}. 
Since the choice of setting 
any of the OPE coefficients equal to zero is arbitrary, one might consider 
a different set, which, at this point, may also seem rather arbitrary:
\begin{eqnarray}
\label{OPEMG}
C_8&=&0.0349\ \mbox{GeV}^{8}\ ,\\
C_{10}&=&-0.0832\ \mbox{GeV}^{10}\ ,\nonumber\\
C_{12}&=&0.161\ \mbox{GeV}^{12}\ ,\nonumber\\
C_{14}&=&-0.17\ \mbox{GeV}^{14}\ ,\nonumber\\
C_{16}&=&-0.55\ \mbox{GeV}^{16}\ .\nonumber
\end{eqnarray}
Even if arbitrary, this choice for $C_D$ with $8\le D\le 16$ is a 
reasonable one.   The values are of the order or magnitude one might 
expect in QCD, with its typical hadronic scale of about 1~GeV. Measured 
in units of 1~GeV, they increase with $D$, but also this is not excluded 
or unnatural, if indeed the OPE is an asymptotic series 
(\seef\ Sec.~\ref{theory}).

Redoing the fits of Tables~\ref{tab1} to \ref{tab4}, but now with 
Eq.~(\ref{OPEMG}) as input, we find the results presented in 
Tables~\ref{tab7} to \ref{tab10}.
\begin{table}[h!]
\hspace{0cm}\begin{tabular}{|c|c|c|c|c|c|}
\hline
& $\a_s(m_\t^2)$ & $C_{4}$ (GeV$^4$) & $C_{6}$ (GeV$^6$)& $C_{8}$ (GeV$^8$)& $\c^2/$dof \\
\hline
FOPT & 0.295(3) & 0.0043(3) & -0.0128(3) & 0.0355(3) & 0.99/1 \\
\hline
CIPT & 0.308(4) & 0.0031(3)  & -0.0129(3) & 0.0354(3) & 0.74/1 \\
\hline
\end{tabular}
\floatcaption{tab7}{{\it Fits as in Table~\ref{tab1},
but with $C_{10}$, $C_{12}$, $C_{14}$ and $C_{16}$ as given in
Eq.~(\ref{OPEMG}).
Errors are statistical only.}}
\end{table}%
\begin{table}[h!]
\hspace{0cm}\begin{tabular}{|c|c|c|c|c|c|}
\hline
& $\a_s(m_\t^2)$ & $C_{4}$ (GeV$^4$) & $C_{6}$ (GeV$^6$)& $C_{8}$ (GeV$^8$)& $\c^2/$dof \\
\hline
FOPT & 0.296(3) & 0.0042(2) & -0.0127(2) & 0.0352(2) & 0.84/1 \\
\hline
CIPT & 0.309(4) & 0.0030(3)  & -0.0128(2) & 0.0351(2) & 0.60/1 \\
\hline
\end{tabular}
\floatcaption{tab8}{{\it Fits as in Table~\ref{tab2},
but with $C_{10}$, $C_{12}$ and $C_{14}$ as given in
Eq.~(\ref{OPEMG}).
Errors are statistical only.}}
\end{table}%
\begin{table}[h!]
\hspace{0cm}\begin{tabular}{|c|c|c|c|c|c|}
\hline
& $\a_s(m_\t^2)$  & $C_{6}$ (GeV$^6$)& $C_{8}$ (GeV$^8$) & $C_{10}$ (GeV$^{10}$)& $\c^2/$dof \\
\hline
FOPT & 0.295(4)  & -0.0130(4) & 0.0356(5) & -0.0836(3) & 1.09/1 \\
\hline
CIPT & 0.308(5)   & -0.0130(4) & 0.0355(5) & -0.0836(3) &  0.84/1 \\
\hline
\end{tabular}
\floatcaption{tab9}{{\it Fits as in Table~\ref{tab3},
but with $C_{12}$, $C_{14}$ and $C_{16}$ as given in
Eq.~(\ref{OPEMG}).
Errors are statistical only.}}
\end{table}%
\begin{table}[h!]
\hspace{0cm}\begin{tabular}{|c|c|c|c|c|c|}
\hline
& $\a_s(m_\t^2)$  & $C_{4}$ (GeV$^4$)& $C_{6}$ (GeV$^6$) & $\c^2/$dof \\
\hline
FOPT & 0.308(8) & 0.0023(12) & -0.009(2)  & 1.73/1 \\
\hline
CIPT & 0.322(11) & 0.0009(15)  & -0.010(2)  & 1.63/1 \\
\hline
\end{tabular}
\floatcaption{tab10}{{\it Fits as in Table~\ref{tab4},
but with $C_{8}$ as given in
Eq.~(\ref{OPEMG}).
Errors are statistical only.}}
\end{table}%

The fits of Tables~\ref{tab7} to \ref{tab9}
are all very good fits, as measured by their $\c^2$ values, 
certainly at least as good as those of Tables~\ref{tab1} to 
\ref{tab3}.\footnote{Table~\ref{tab10} shows some tension for the $C_{4,6}$ 
coefficients relative to the values shown in Tables~\ref{tab7} to \ref{tab10}. 
This is because DVs have not yet been taken into account, as will be seen in 
Table~\ref{tab14} below.  Recall that the set~(\ref{lesspinched}) contains
two weights which are not doubly or triply pinched.} The implication of this exercise is that 
there could be at least two solutions, depending on what one assumes 
for the values of higher-dimension OPE coefficients. The existence of these 
two (and possibly more) solutions reveals a fundamental problem of the 
truncated-OPE strategy:  the solution found by this strategy depends
on the choice of the values of the OPE coefficients not included in the 
fits. In addition, the strategy does not provide a physics argument, 
either {\it a priori} or {\it a posteriori}, for what choice to make. The
solution with Eq.~(\ref{OPEMG}) as input leads to values of $\a_s(m_\t^2)$ that 
are about 0.025, or 8\%, lower than those obtained using the alternate
input set in which the relevant higher-dimension $C_D$ are set to zero by 
hand. We note that the values for $C_{10}$ in Table~\ref{tab9}, 
and $C_8$ in Tables~\ref{tab7} to \ref{tab9}, for which these coefficients are
not input, agree well with the value in Eq.~(\ref{OPEMG}). Likewise, if 
one explores tests like those of Table~\ref{tab5}, the results are 
internally consistent as well as consistent with the values in Eq.~(\ref{OPEMG}).
Without further information, it is not possible to claim that one solution 
is better than the other.  To summarize: the internal consistency 
among all fits cannot be used as a reliable test for judging the robustness of
the result for $\a_s(m_\t^2)$, in contrast to what is advocated in 
Ref.~\cite{Pich}, because the solution described in this subsection
passes all the same consistency tests.

We also considered the ``secondary'' tests of Tables~5 and 9 of 
Ref.~\cite{Pich}.\footnote{For Table~9, see Sec.~\ref{DVs} below.}
In Table~5 of Ref.~\cite{Pich} the twelve moments with weights~(\ref{optimal}) 
were employed choosing $m=0,\ 1$, $n=0,\dots,5$, and $s_0=2.8$~GeV$^2$. 
For each moment a value of $\a_s(m_\t^2)$ was extracted ignoring all 
non-perturbative contributions, \ie, setting all $C_{D\ge 2}=0$ and 
ignoring DVs. While the results appeared to suggest self-consistency 
and consistency with all other fits, we found that this is only the case 
because Ref.~\cite{Pich} limits itself to the 
single choice $s_0=2.8$~GeV$^2$.

Since $s_0=2.6$~GeV$^2$ corresponds to the bin immediately before 
$s_0=2.8$~GeV$^2$, we have varied $s_0$ in the range $s_0=2.6$~GeV$^2$ to 
$s_0=m_\t^2$, and considered the differences in the values of $\a_s(m_\t^2)$ 
obtained from these twelve moments. In computing these differences, it is 
important to take correlations into account, since the integrated data, 
and thus the fits, are highly correlated. Of course, such differences 
should be consistent with zero, within errors. Instead, we find that 
these differences are often inconsistent with zero at the 2 to 4 $\s$ level, 
depending on which pair of moments one considers. 
Thus, instead of confirming the robustness claimed in Ref.~\cite{Pich}, 
the results obtained using these moments actually point 
to potential internal inconsistency problems.

Since some of these moments are only singly pinched, and since also all 
$C_{D\ge 2}$ were set equal to zero, one might argue that it is these
shortcomings which are the source of the non-zero differences noted above.
Even if this is the case, the conclusion remains that the results of 
Table~5 of Ref.~\cite{Pich} cannot be taken as providing any additional 
evidence for the validity of the truncated-OPE strategy.

\subsubsection{\label{DVs} The omission of duality violations}

We now turn to the second assumption made in the truncated-OPE-model strategy.
While the previous subsection revealed a major problem with the truncation of 
the OPE itself, one might still think that the use of weights that are at 
least doubly pinched makes it safe to ignore DVs. Before we carry out another 
exercise to probe this assumption quantitatively, let us consider this 
assumption in the light of the data.
\begin{figure}[t!]
\begin{center}
\includegraphics*[width=12cm]{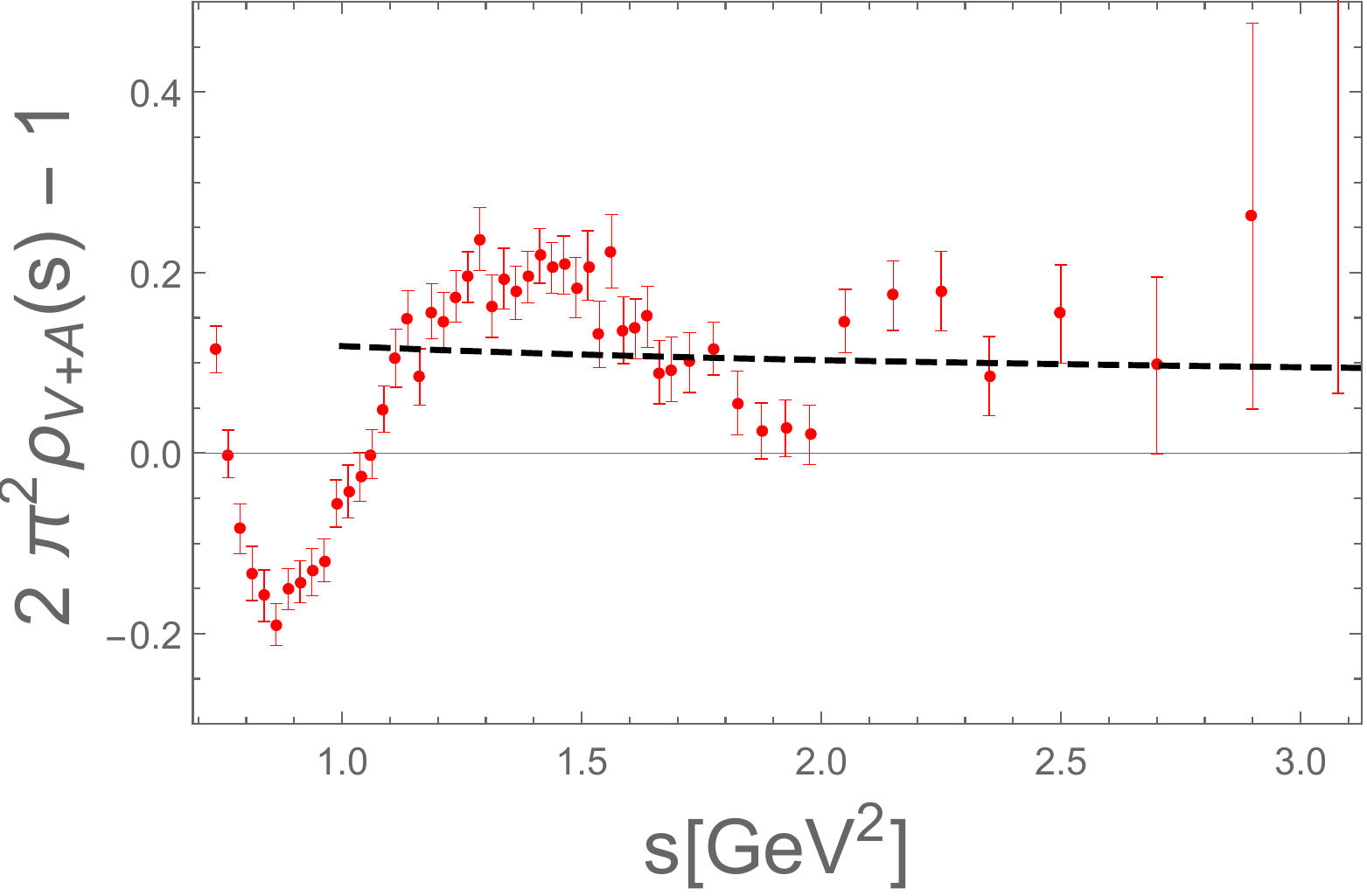}
\end{center}
\begin{quotation}
\vspace*{-2ex}
\floatcaption{V+A-1}%
{\it Blow-up of the ALEPH data (red data points) in the
large-$s$ region of the $V+A$ non-strange spectral function
($s$ in {\rm GeV}$^2$). What is shown is $2\p^2\r_{V+A}-1$, \ie,
the dynamical QCD contribution to the spectral distribution.
Black dashed line:  perturbation theory (CIPT) representation of
the model, also with the parton-model contribution subtracted.}
\end{quotation}
\vspace*{-4ex}
\end{figure}
In Fig.~\ref{V+A-1} we show the large-$s$ region of the non-strange, $V+A$
spectral-function obtained from ALEPH data \cite{ALEPH13}. 
We have plotted $2\p^2\r_{V+A}-1$, rather than
$2\p^2\r_{V+A}$, in order to remove the free-quark (or parton-model) 
contribution, which is independent of QCD dynamics; $2\p^2\r_{V+A}-1$
thus represents the dynamical QCD contribution to the spectral distribution.
The difference between the dashed curve and 
the horizontal axis in Fig.~\ref{V+A-1} represents the perturbative
part of the dynamics from which $\a_s$ is extracted.
It is clear that DVs, represented by the oscillations of the data around the
dashed curve, are not a small part of the dynamical QCD contribution
to the spectral function in this region. This is especially evident
in the region below $s=2.3$~GeV$^2$, where the data errors are small. 
In addition, there is no sign in this region of a strong damping of 
these oscillations. Therefore, even though above $s=2.3$~GeV$^2$ the 
errors are large enough for the data to be in rough agreement
with the dashed curve, it is not safe to assume that DVs are 
small enough to be irrelevant in the region up to $s=m_\t^2$.

We can study the effects of DVs quantitatively. As a first exercise, we 
consider how a quantitatively reasonable representation of the 
duality-violating part of the spectral function, 
$\r_{V+A}^{\rm DV}(s)=\r_V^{\rm DV}(s)+\r_A^{\rm DV}(s)$, affects the 
results of the truncated-OPE strategy. We take our representation of DVs 
from one of the fits of Ref.~\cite{alphas14}, and assume values for the 
OPE coefficients $C_{10}$ through $C_{16}$ which are consistent with 
FESRs that account for these DVs. From Ref.~\cite{alphas14},
we take\footnote{In more detail, we take these values from Table~5 
of Ref.~\cite{alphas14}, $s_{\rm min}=1.55$~GeV$^2$, CIPT. The FOPT values 
are the same within errors.}
\begin{eqnarray}
\label{DVparvalues}
\d_V&=3.35 \ ,\qquad\qquad\d_A&=1.59\ ,\\
\g_V&=0.70 \ ,\qquad\qquad\g_A&=1.44\ ,\nonumber\\
\a_V&=4.00 \ ,\qquad\qquad\a_A&=5.37 \ ,\nonumber\\
\b_V&=4.23 \ ,\qquad\qquad\b_A&=2.03 \ ,\nonumber
\end{eqnarray}
(with $\g_{V,A}$ and $\b_{V,A}$ in GeV$^{-2}$).
In Sec.~7 of Ref.~\cite{alphas14} we used the results of this fit to 
estimate the values for all OPE coefficients $C_D$, $D=4,\dots,16$.
For the CIPT case, the estimates we need here are
\begin{eqnarray}
\label{OPE74}
C_8&=&0.0349~\mbox{GeV}^8\ ,\\
C_{10}&=&-0.0832~\mbox{GeV}^{10}\ ,\nonumber\\
C_{12}&=&0.161~\mbox{GeV}^{12}\ ,\nonumber\\
C_{14}&=&-0.191~\mbox{GeV}^{14}\ ,\nonumber\\
C_{16}&=&-0.233~\mbox{GeV}^{16}\ .\nonumber
\end{eqnarray}
For the exercise below, whose purpose is to illustrate the sensitivity of
the output $\alpha_s$ to the input values assumed for the higher
dimension $C_D$, it suffices to use these same values for the FOPT
exploration as well.

Assuming the values given in Eq.~(\ref{OPE74})
for $C_8$ through $C_{16}$, and keeping the second
term on the right-hand side of Eq.~(\ref{sumrule}), using the DV parameters
of Eq.~(\ref{DVparvalues}), we find the results shown in Tables~\ref{tab11}, 
\ref{tab12}, \ref{tab13} and \ref{tab14} by applying the 
strategy of Ref.~\cite{Pich} using the weights in Eq.~(\ref{ALEPH}), the 
reduced weights~(\ref{reduced}),
the ``optimal'' weights~(\ref{optimal}),
or the weights~(\ref{lesspinched}).
\begin{table}[h!]
\hspace{0cm}\begin{tabular}{|c|c|c|c|c|c|}
\hline
& $\a_s(m_\t^2)$ & $C_{4}$ (GeV$^4$) & $C_{6}$ (GeV$^6$)& $C_{8}$ (GeV$^8$)& $\c^2/$dof \\
\hline
FOPT & 0.297(3) & 0.0042(3) & -0.0126(3) & 0.0353(3) & 1.30/1 \\
\hline
CIPT & 0.310(4) & 0.0029(4)  & -0.0124(3) & 0.0352(3) & 1.00/1 \\
\hline
\end{tabular}
\floatcaption{tab11}{{\it Fits as in Table~\ref{tab1},
but with $C_{10}$, $C_{12}$, $C_{14}$ and $C_{16}$ as given in
Eq.~(\ref{OPE74}), and including the DV parameters of Eq.~(\ref{DVparvalues}).
Errors are statistical only.}}
\end{table}%

\begin{table}[h!]
\hspace{0cm}\begin{tabular}{|c|c|c|c|c|c|}
\hline
& $\a_s(m_\t^2)$ & $C_{4}$ (GeV$^4$) & $C_{6}$ (GeV$^6$)& $C_{8}$ (GeV$^8$)& $\c^2/$dof \\
\hline
FOPT & 0.297(3) & 0.0041(2) & -0.0126(2) & 0.0352(2) & 1.20/1 \\
\hline
CIPT & 0.310(3) & 0.0028(2)  & -0.0126(2) & 0.0351(1) & 0.90/1 \\
\hline
\end{tabular}
\floatcaption{tab12}{{\it Fits as in Table~\ref{tab2},
but with $C_{10}$, $C_{12}$ and $C_{14}$ as given in
Eq.~(\ref{OPE74}), and including the DV parameters of Eq.~(\ref{DVparvalues}).
Errors are statistical only.}}
\end{table}%

\begin{table}[h!]
\hspace{0cm}\begin{tabular}{|c|c|c|c|c|c|}
\hline
& $\a_s(m_\t^2)$  & $C_{6}$ (GeV$^6$)& $C_{8}$ (GeV$^8$) & $C_{10}$ (GeV$^{10}$)& $\c^2/$dof \\
\hline
FOPT & 0.296(4)  & -0.0127(4) & 0.0354(5) & -0.0834(3) & 1.36/1 \\
\hline
CIPT & 0.310(5)   & -0.0128(4) & 0.0353(5) & -0.0834(3) &  1.06/1 \\
\hline
\end{tabular}
\floatcaption{tab13}{{\it Fits as in Table~\ref{tab3},
but with $C_{12}$, $C_{14}$ and $C_{16}$ as given in
Eq.~(\ref{OPE74}), and including the DV parameters of Eq.~(\ref{DVparvalues}).
Errors are statistical only.}}
\end{table}%
\begin{table}[h!]
\hspace{0cm}\begin{tabular}{|c|c|c|c|c|c}
\hline
& $\a_s(m_\t^2)$ & $C_{4}$ (GeV$^4$) & $C_{6}$ (GeV$^6$)& $\c^2/$dof \\
\hline
FOPT & 0.301(9) & 0.004(1) & -0.012(2)  & 2.04/1 \\
\hline
CIPT & 0.313(11) & 0.003(1)  & -0.012(2)  & 1.95/1 \\
\hline
\end{tabular}
\floatcaption{tab14}{{\it Fits as in Table~\ref{tab4},
but with $C_{8}$ as given in
Eq.~(\ref{OPE74}), and including the DV parameters of Eq.~(\ref{DVparvalues}).
Errors are statistical only.}}
\end{table}%
Again, these fits are good fits, and they are consistent with each other.
We note that the fits of Table~\ref{tab14} are
more susceptible to DVs, because the weights~(\ref{lesspinched}) include
polynomials which are less pinched than the weights of the other sets.
We also note that including DVs changes the results of Table~\ref{tab10} into 
those shown in Table~\ref{tab14}, which are in excellent agreement 
with the results in the other tables.
The FOPT values for $\a_s(m_\t^2)$ are about 0.02 lower than those in
Tables~\ref{tab1}, \ref{tab2}, \ref{tab3} and \ref{tab4}, while the CIPT 
values are about 0.025 lower. Perhaps not surprisingly, they are in 
good agreement with the values found in Ref.~\cite{alphas14}. This suggests 
that the OPE can possibly be trusted at $s\approx -m_\t^2$ up to $D=16$, 
even if it is an asymptotic expansion. However, it is also clear that 
solutions to the truncated-OPE fit strategy exist with OPE coefficients 
that cannot be considered small enough to be set equal to zero beyond 
$C_8$ (or $C_{10}$ for fits with ``optimal'' weights), and with DVs 
that cannot be neglected.

We also redid the ``$A^{\o^{(21)}}$'' fits of Table~9 of Ref.~\cite{Pich}.
What is done here is to take the moment with $w_{00}$ of Eq.~(\ref{ALEPH}) 
for values of $s_0$ ranging from $s_0=2.0$~GeV$^2$ to $s_0=m_\t^2$, and 
fit these 9 data points as a function of $s_0$ to a fit function that 
includes $\a_s(m_\t^2)$, $C_6$ and $C_8$, ignoring DVs.
We reproduced the values of Table~9 of Ref.~\cite{Pich}. Redoing these fits
with DVs, we find values consistent with those of Tables~\ref{tab11} to 
\ref{tab14}, instead of Tables~\ref{tab1} to \ref{tab4}. Again,
we conclude that the test of Table~9 of Ref.~\cite{Pich}
does not provide a proof of the stability of the results, in contrast to
what is suggested in Ref.~\cite{Pich}.

Finally, Ref.~\cite{Pich} introduces yet another set of weights that have an 
additional exponential 
suppression, similar in spirit to the moments employed in the SVZ sum rules 
of Ref.~\cite{SVZ}. Specifically, Ref.~\cite{Pich} considers a set of moments with 
weights $w_B(a,n)= (1-x^{n+1})e^{-a x}$, with $a\ge 0$. This type of 
moment acquires contributions from OPE condensates of all dimensions. 
In Ref.~\cite{Pich}, $\alpha_s$ was extracted from a single sum rule at a time, 
ignoring all non-perturbative corrections, for several values of $s_0$ and 
the Borel parameter $a$. We have reproduced their results\footnote{We
restricted ourselves to CIPT.} and, in comparison 
to the plots of Ref.~\cite{Pich}, we find numerical agreement. The stability of 
the results regarding non-perturbative physics can be investigated by
adding, successively, higher order terms in the OPE as well as
adding or removing the DV contribution to the moments.  For this
exercise we employed the condensates of Eq.~(\ref{OPE74}), as well as
\begin{eqnarray}
\label{OPEmodel}
C_4&=&0.00268  \ \mbox{GeV}^4\ ,\\
C_6&=&-0.0125  \ \mbox{GeV}^6\ .\nonumber
\end{eqnarray}
These values are again taken from Sec.~7 of Ref.~\cite{alphas14}.  The
results thus obtained for $\alpha_s$ start stabilizing with respect
to the OPE only after the term with $D= 14$ is included. Together with
the addition of the DV contribution, the results for $\alpha_s$
become then fully consistent with those of Tabs.~\ref{tab11},
\ref{tab12}, \ref{tab13} and \ref{tab14}.  Values for $\alpha_s$ are
systematically lower than in Ref.~\cite{Pich} and in good agreement
with the ones found in Ref.~\cite{alphas14}. In addition, the remaining
instability with respect to the Borel parameter $a$
observed in Ref.~\cite{Pich} (for CIPT) is eliminated when the
non-perturbative contributions are properly taken into account.
We conclude that also this exploration does not validate the solution
claimed by Ref.~\cite{Pich}.

\bigskip
In this section, we found that it is easy to find solutions to fits based 
on the truncated-OPE strategy yielding significantly different values for 
$\a_s(m_\t^2)$. Of course, at this point, none of these explorations tells 
us which solution is closest to the truth. Maybe none of them is; based 
on the exercises in this section, we cannot exclude the existence of yet 
other solutions to this collection of fits based on the 
truncated-OPE strategy. Even if the solution found in Ref.~\cite{Pich} would be
the correct one, our results imply that it is impossible to assign a 
reliable systematic error to the values found for $\a_s(m_\t^2)$ based 
on the truncated-OPE-model strategy. 

However, a very different type of test can be performed, in which we 
consider data constructed from a model compatible with the experimental
V+A spectral function and having a known value for $\a_s(m_\t^2)$ and known 
DV contributions. The question then becomes whether the truncated-OPE 
strategy, applied to data constructed using this model, is able to 
successfully reproduce the known value of $\a_s(m_\t^2)$.
We describe such a test in the next section.

\section{\label{fake data} A numerical experiment}
In this section, we will carry out the ``fake data'' test. We start from a 
model of the $V+A$ spectral function which gives a good description of the 
real data from $s=1.55$~GeV$^2$ to $m_\t^2$. The model value for the strong 
coupling is taken to be $\a_s(m_\t^2)=0.312$ (using CIPT), and the model 
has non-negligible DVs compatible with the real data; the corresponding 
parameters are given in Eq.~(\ref{DVparvalues}). A multivariate Gaussian 
distribution is defined with model values at the ALEPH bin energies as 
central values, and with fluctuations around it controlled by the 
real-data covariance matrix \cite{ALEPH13}. This distribution is used to 
generate, probabilistically, a fake data set. To this fake data set we 
apply the truncated-OPE-model fits. We assess the reliability of the
truncated-OPE-model by comparing the resulting fit values for $\a_s(m_\t^2)$ 
to the underlying true model value.

The key point is the following. Essentially, Ref.~\cite{Pich} claims that
it is not necessary to take DVs explicitly into account, \ie, that they can 
be neglected for fits involving the (typically at least doubly-pinched)
weights employed in previous implementations of the truncated-OPE strategy,
\seef\ Sec.~\ref{PRresults}.
For the model, we know the value of $\a_s(m_\t^2)$ explicitly, and also 
know that it has significant DVs, by construction.  It is also
realistic, since it describes the spectral function data very well.
Therefore, the truncated-OPE strategy, if reliable, should recover the 
model value of $\a_s(m_\t^2)$. If it does, the truncated-OPE strategy would 
pass this non-trivial test. If it does not, \ie, if it fails to recover 
the model value of $\a_s(m_\t^2)$ within statistical errors, the implication 
is that this strategy is incapable, in general, of finding the correct 
value from the real data with meaningful errors and is, thus, unreliable.

\begin{figure}
\begin{center}
\includegraphics*[width=13cm]{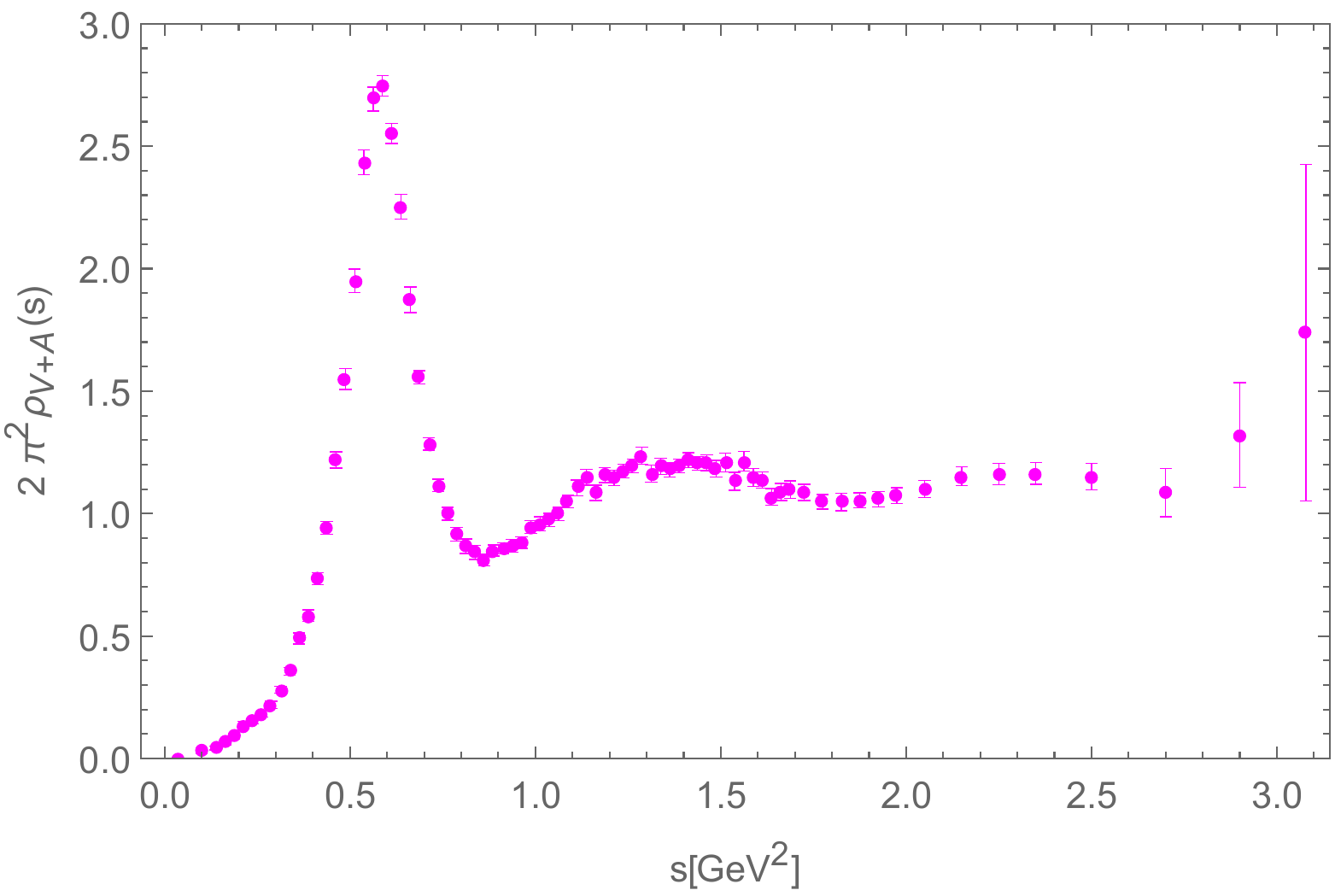}
\vspace{0.0cm}
\includegraphics*[width=13cm]{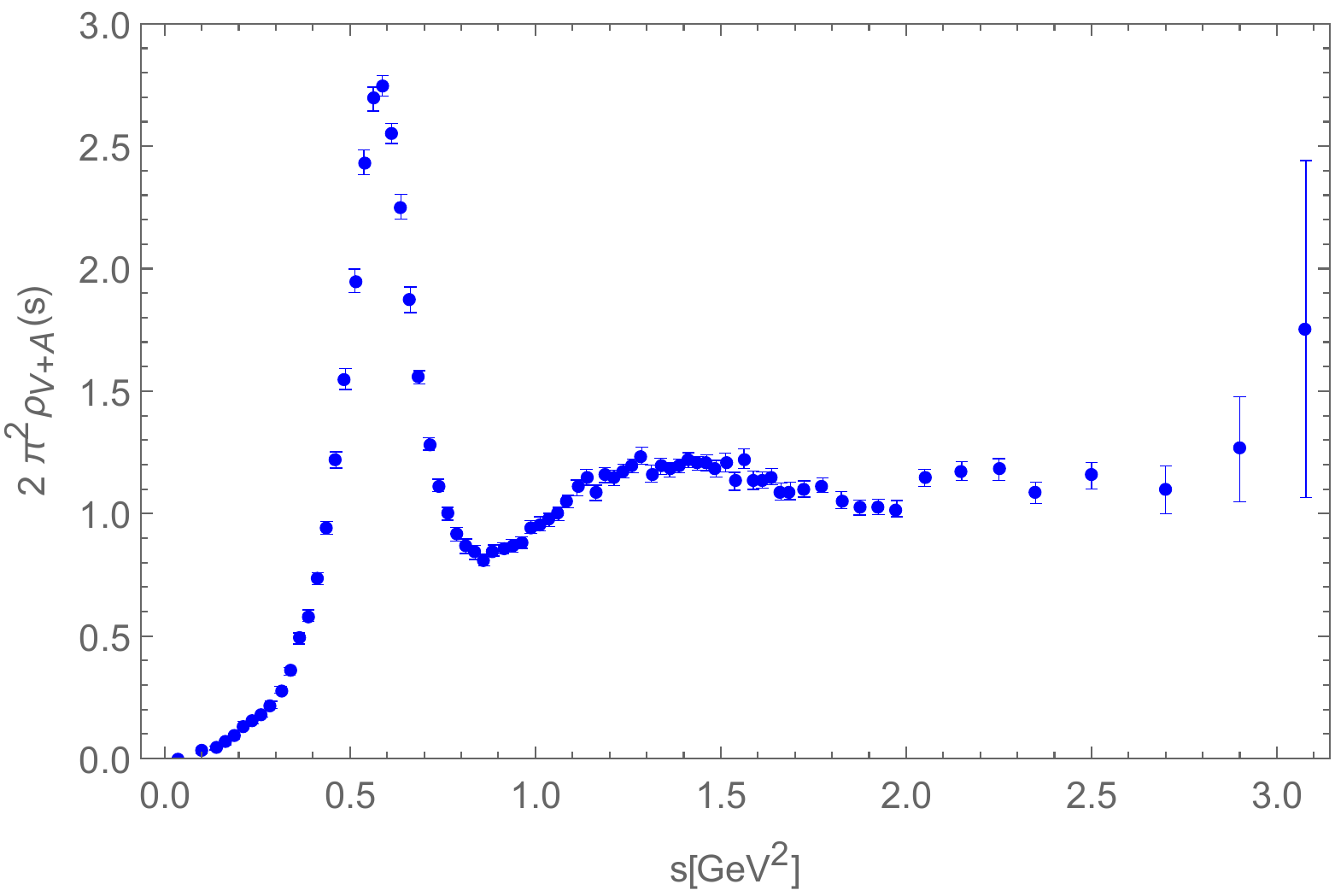}
\end{center}
\begin{quotation}
\floatcaption{fake}%
{\it $V+A$ non-strange spectral function. Top panel: fake data, generated
as described in the text, as a function of $s$. Bottom panel:
true ALEPH data \cite{ALEPH13} as a function of $s$.
The fake data have been generated for $s\ge 1.55$~{\rm GeV}$^2$; below this
value the two data sets are the same.}
\end{quotation}
\vspace*{-4ex}
\end{figure}
\begin{figure}
\begin{center}
\includegraphics*[width=7.4cm]{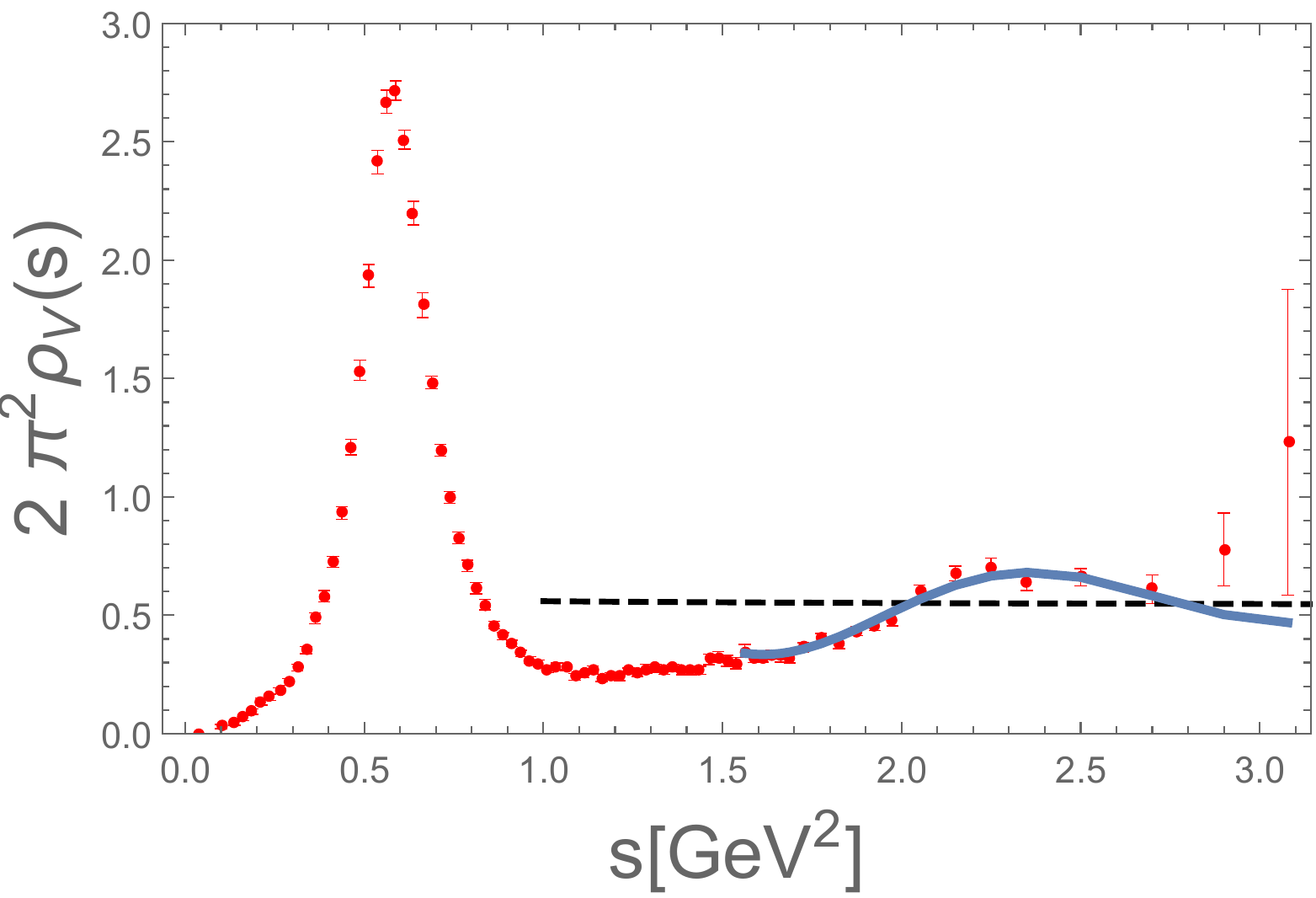}
\hspace{0.0cm}
\includegraphics*[width=7.4cm]{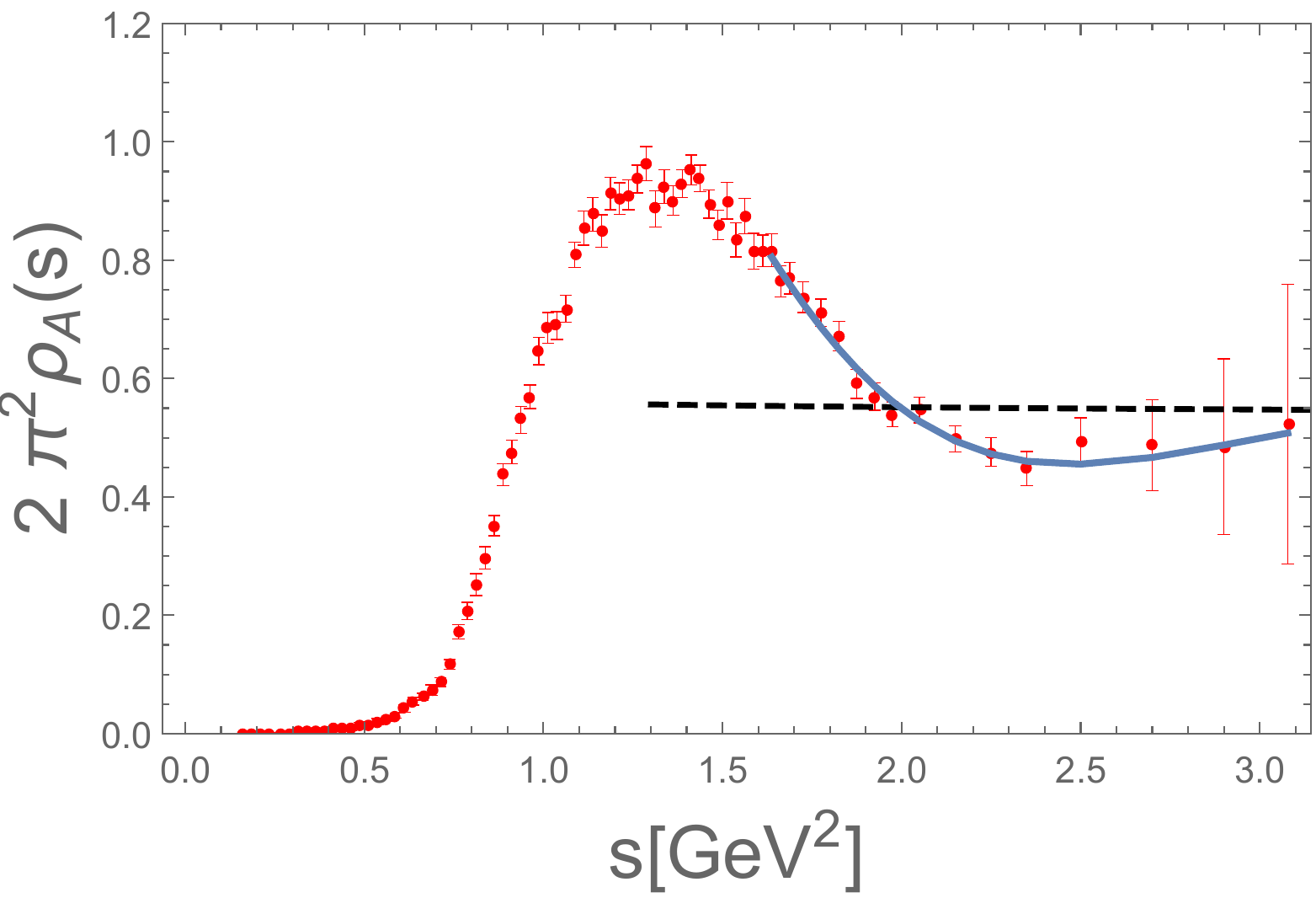}
\end{center}
\begin{quotation}
\floatcaption{VandA}%
{\it $V$ (left panel) and $A$ (right panel) non-strange spectral functions, 
as a function of $s$. Data points from Ref.~\cite{ALEPH13}, curves represent 
the model described in the text for $s\geq 1.55\, \mathrm{GeV}^2$, and dashed lines show the perturbative 
part of the model.   }
\end{quotation}
\end{figure}

\subsection{\label{generation} Fake data}
We show the real data and the fake data in Fig.~\ref{fake}.   These fake data
have been generated by a model using CIPT for the perturbative part with
$\a_s(m_\t^2)=0.312$.   Correspondingly, we will carry out our test using
CIPT.\footnote{Similar tests can be carried out with an FOPT-based fake
data set and FOPT fits, with very similar results.}
Note that the fake data resemble the real data strongly, and that,
by construction, the covariance matrices for both data sets are identical.%
\footnote{It appears that the fake data set is slightly smoother than the 
real data set.} The DV parameters defining the duality-violating part of 
the fake data are given in Eq.~(\ref{DVparvalues}). The ALEPH $V$ and $A$
spectral functions, together with the model representations using these DV
parameters, are shown in Fig.~\ref{VandA}. The model OPE coefficients 
follow from the exact FESR~(7.3) in Ref.~\cite{alphas14}, and have been 
given already in Eqs.~(\ref{OPE74}) and~(\ref{OPEmodel}).

\subsection{\label{test} Test of the truncated-OPE-model strategy}
We apply the truncated-OPE-model strategy directly to the fake data.
Tables~\ref{tab15} to \ref{tab18} employ the same fits used to 
produce Tables~\ref{tab1} to \ref{tab4}, except that now the real data 
have been replaced by the fake data. We show only CIPT fit results 
because the fake data have been generated from a model based on the 
CIPT perturbative scheme. This is, however, not essential; 
the same exercise can also be carried out for FOPT.

We see that the truncated-OPE strategy fails to reproduce the model value for
$\a_s(m_\t^2)=0.312$ (by 5 to 7 $\s$ for Tables~\ref{tab15} to 
\ref{tab17}), even though the individual fits have good $\chi^2$ values, 
and results of the different fits look mutually consistent.
The same is true of the results for the OPE coefficients, 
which come out much smaller in magnitude than the values given in 
Eqs.~(\ref{OPE74}) and~(\ref{OPEmodel}).  
In addition to this failure to reproduce the 
model parameter values, the results of this exercise also once more 
show that demonstrating internal consistency among the various fits of 
Ref.~\cite{Pich} does not allow one to conclude that the determination of $\a_s(m_\t^2)$ employing the truncated-OPE 
strategy is valid within its quoted
errors.  We have verified that the correct values of $\a_s(m_\t^2)$
and the OPE coefficients are reproduced, within statistical errors, 
if higher-dimension OPE coefficients and DVs are used as input for 
the fits, analogous to the tests in Tables~\ref{tab11} to \ref{tab14}.
This exercise shows that not only does the truncated-OPE strategy not
distinguish between significantly different solutions, but that,
in general, its assumptions may end up driving it to a ``solution''
which is actually incorrect.

\begin{table}[h!]
\hspace{0cm}\begin{tabular}{|c|c|c|c|c|}
\hline
$\a_s(m_\t^2)$ & $C_{4}$ (GeV$^4$) & $C_{6}$ (GeV$^6$)& $C_{8}$ (GeV$^8$)& $\c^2/$dof \\
\hline
0.334(4) & -0.0023(4)  & 0.0007(3) & -0.0008(4) & 0.94/1 \\
\hline
\end{tabular}
\floatcaption{tab15}{{\it CIPT fits employing the truncated-OPE strategy 
on the fake data, based on the weights of Eq.~(\ref{ALEPH}).
By assumption, $C_{10}=C_{12}=C_{14}=C_{16}=0$.
Errors are statistical only.}}
\end{table}%
\begin{table}[h!]
\hspace{0cm}\begin{tabular}{|c|c|c|c|c|}
\hline
 $\a_s(m_\t^2)$ & $C_{4}$ (GeV$^4$) & $C_{6}$ (GeV$^6$)& $C_{8}$ (GeV$^8$)& $\c^2/$dof \\
\hline
 0.334(3) & -0.0023(3)  & 0.0007(2) & -0.0007(2) & 0.98/1 \\
\hline
\end{tabular}
\floatcaption{tab16}{{\it CIPT fits employing the truncated-OPE strategy 
on the fake data, based on the reduced weights~(\ref{reduced}). By assumption, 
$C_{10}=C_{12}=C_{14}=0$. Errors are statistical only.}}
\end{table}%
\begin{table}[h!]
\hspace{0cm}\begin{tabular}{|c|c|c|c|c|}
\hline
 $\a_s(m_\t^2)$  & $C_{6}$ (GeV$^6$)& $C_{8}$ (GeV$^8$) & $C_{10}$ (GeV$^{10}$)& $\c^2/$dof \\
\hline
 0.334(4)   & 0.0008(4) & -0.0008(5) & 0.0001(3) &  0.92/1 \\
\hline
\end{tabular}
\floatcaption{tab17}{{\it CIPT fits employing the truncated-OPE strategy 
on the fake data, based on the ``optimal'' weights~(\ref{optimal}) with 
$m=1$ and $n=1,\dots\,5$. By assumption, $C_{12}=C_{14}=C_{16}=0$.
Errors are statistical only.}}
\end{table}%
\begin{table}[h!]
\hspace{0cm}\begin{tabular}{|c|c|c|c|c|}
\hline
 $\a_s(m_\t^2)$  & $C_{4}$ (GeV$^4$)& $C_{6}$ (GeV$^6$) & $\c^2/$dof \\
\hline
 0.337(11) & -0.003(2)  & 0.001(2)  & 1.25/1 \\
\hline
\end{tabular}
\floatcaption{tab18}{{\it CIPT fits employing the truncated-OPE strategy 
on the fake data, based on the weights of Eq.~(\ref{lesspinched}) with 
$n=0,\dots\,3$. By assumption, $C_{8}=0$. Errors are statistical only.}}
\end{table}%

\newpage
\subsection{\label{discussion4} Discussion}
It is instructive to ask why the truncated-OPE-model strategy fails to 
reproduce the model values of $\a_s(m_\t^2)$ and the OPE coefficients.
As we have seen in Secs.~\ref{truncation} and \ref{DVs}, setting the high-dimension OPE coefficients and the DVs to zero affects significantly the value of  
$\a_s(m_\t^2)$ extracted from the fits.  Here, since we have the explicit spectral function for the fake data in hand, we can analyze the effects of the known DVs
underlying these data on the fit strategy.   These DVs affect the extraction of $\a_s$ directly through the term on the right hand side of Eq.~(\ref{sumrule}), and also indirectly 
through the non-zero values shown in 
Eqs.~(\ref{OPE74}) and~(\ref{OPEmodel}) of the OPE condensates, 
$C_D$, obtained through Eq. (7.3) of 
Ref.~\cite{alphas14}.

Let us consider Fig.~\ref{DVVplusA}, which shows again a blow-up of 
the large-$s$ region of the $V+A$ spectral function. The red experimental 
points represent the ALEPH data, and the thick blue curve shows the model 
representation of the $V+A$ spectral function, which is the sum of the 
model representations of the $V$ and $A$ spectral functions shown in 
Fig.~\ref{VandA} and as the blue dot-dashed curves in Fig.~\ref{DVVplusA}. The black 
dashed curve shows the perturbative part of the V+A model representation.

There are several important observations to make about this figure. First, we
note that the model is not excluded by the data, even if one can imagine
other models that might do equally well. Second, let us reiterate that
it is not correct to think of DVs in this region of the spectral function 
as a ``small effect.'' The parton model (\ie, QCD to zeroth order in $\a_s$) 
contribution is given by a horizontal line at $2 \pi^2\r_{V+A}=1$.
As already emphasized in Sec.~\ref{DVs},
it is the difference between the actual spectral function and 
this parton model horizontal line that contains the dynamics of 
QCD, and the duality-violating oscillations are not small on this 
scale. Third, one notes that the blue curve shows a duality-violating 
oscillation that is quite large at $s_0=m_\t^2$, larger, in fact,
than at any other value of $s_0$ larger than 1.7~GeV$^2$. This can
happen over a limited range of $s$ even though the individual 
$V$ and $A$ DVs are exponentially damped. Since the truncated-OPE 
strategy of Ref.~\cite{Pich} simply assumes DVs to be suppressed to a negligible level in its $s_0=m_\tau^2$ fits
without being able to test this assumption for validity, the result
is that it is unable to reproduce the model value for $\a_s(m_\t^2)$
correctly.
\begin{figure}[t!]
\begin{center}
\includegraphics*[width=13cm]{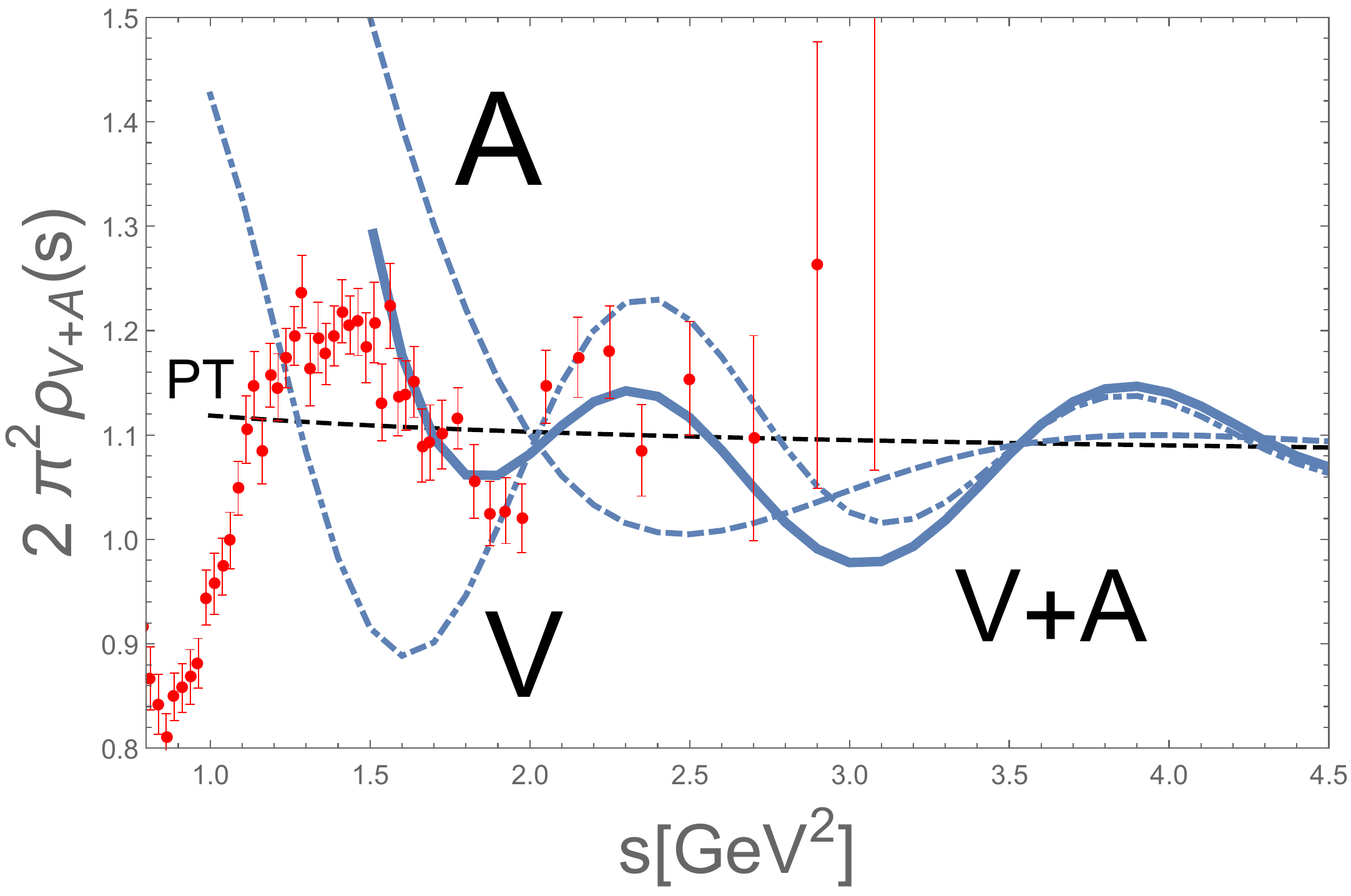}
\end{center}
\begin{quotation}
\vspace*{-2ex}
\floatcaption{DVVplusA}%
{\it Blow-up of the large-$s$ region of the $V+A$ non-strange spectral function.
Black dashed line: the perturbative (CIPT) representation of
the model. Blue curve: full model representation, including DVs. 
Blue dot-dashed curves: separate $V$ and $A$ parts of the model spectral 
function, shown also in Fig.~\ref{VandA}.}
\end{quotation}
\vspace*{-4ex}
\end{figure}

\section{\label{DVdiscussion} Modeling duality violations}
There are two lessons to be learned from the failure of the truncated-OPE-model
approach.   First, DVs cannot simply be ignored.   The data do not exclude the
possibility that they are significant enough that they have to be taken into
account in any high-precision fit of the data, which is, by experimental
necessity, limited to $s\le m_\t^2$. This means, in practice, that DVs have to be
modeled in order to carefully
assess their contribution to any quantity extracted from these data.\footnote{We have carried out very
extensive searches for sets of weights for which DVs contribute 
insignificantly to all associated moments. While we have no
proof that such a set cannot be found, we have not succeeded
in finding one.}

Of course, modeling DVs, using Eq.~(\ref{ansatz}), does two things. First,
it introduces a new assumption into the analysis --- the assumption that the
model is good enough that results obtained for $\a_s(m_\t^2)$ from $\t$
decays are reliable. While clearly the model~(\ref{ansatz}) does a good
job representing the data, it is possible that it does not give an accurate
representation of the $V$ and $A$ spectral functions for $s>m_\t^2$,
where it is needed in the sum rule~(\ref{sumrule}), but where
no data are available \cite{alphas1}. However, ignoring DVs altogether
in any type of fit to the data amounts to setting $\r_{V/A}^{\rm DV}(s)=0$
in Eq.~(\ref{ansatz}). Clearly, this is nothing else than a different choice of
model. Given the oscillations visible in the data, we believe the
choice that ignores DVs, in fact, to be a very poor model. In any case, 
our analysis in Secs.~\ref{PRresults} and \ref{fake data} demonstrates 
that the model in which DVs are ignored does not lead to reliable results, 
{\it irrespective} of the question of the reliability of introducing an explicit 
model for DVs.

Second, once one models DVs explicitly, one is able to avoid
artificially truncating the OPE. Instead, one can use the $s_0$ dependence 
of the spectral integrals in the region where the theoretical 
representation composed of the OPE and the DV {\it ansatz} works, 
thus avoiding spectral integrals involving weights that probe very 
high orders in the OPE. This approach, the DV-model strategy,
was developed in Ref.~\cite{alphas1} and applied there, and in Ref.~\cite{alphas2},
to the OPAL data \cite{OPAL}, and in Ref.~\cite{alphas14} to the revised
ALEPH data \cite{ALEPH13}.

\subsection{\label{DV-model} Summary of the DV-model strategy of Ref.~\cite{alphas14}}
We will not review, in this article, the DV-model strategy 
employed in our analyses of the $\t$-decay data, as
it has been explained in great detail in Refs.~\cite{alphas1,alphas2,alphas14}.
As indicated above, a model of the form~(\ref{ansatz}) was used
to parametrize DVs separately in the $V$ and $A$ channels, 
and the analysis was restricted to FESRs involving weights
that probe OPE coefficients only up to dimension eight. Note that we
also avoided weights with a term linear in $x$, which probe 
$C_4$, because of potential problems with such moments already 
in perturbation theory \cite{alphas1,BBJ12}. We varied 
$s_0\in[s_{\rm min},m_\t^2]$ with $1.4~$GeV$^2\le s_{\rm min}\le 1.7~$GeV$^2$, 
checking for stability as a function of $s_{\rm min}$, and carrying out many 
self-consistency tests between a large number of fits.\footnote{We 
revisit one such stability test in Sec.~\ref{criticism} below.}

One of the tests we carried out is to consider the $s_0$ dependence of
our fitted representation in comparison with the data for the spectral
integrals with moments~(\ref{ALEPH}). We show, in Fig.~\ref{wklus},
the results based on our $s_{\rm min}=1.55$~GeV$^2$ CIPT fit in Table~5 of 
Ref.~\cite{alphas14}. What is plotted in each figure is the $s_0$-dependence
of the spectral integral at $s_0=m_\t^2$ minus the spectral integral at 
$s_0$. The presence of strong correlations in the data and the fits makes 
it necessary to plot such differences if one wishes to appropriately
appraise the level of agreement between theory and data.
\begin{figure}[t!]
\begin{center}
\includegraphics*[width=7.9cm]{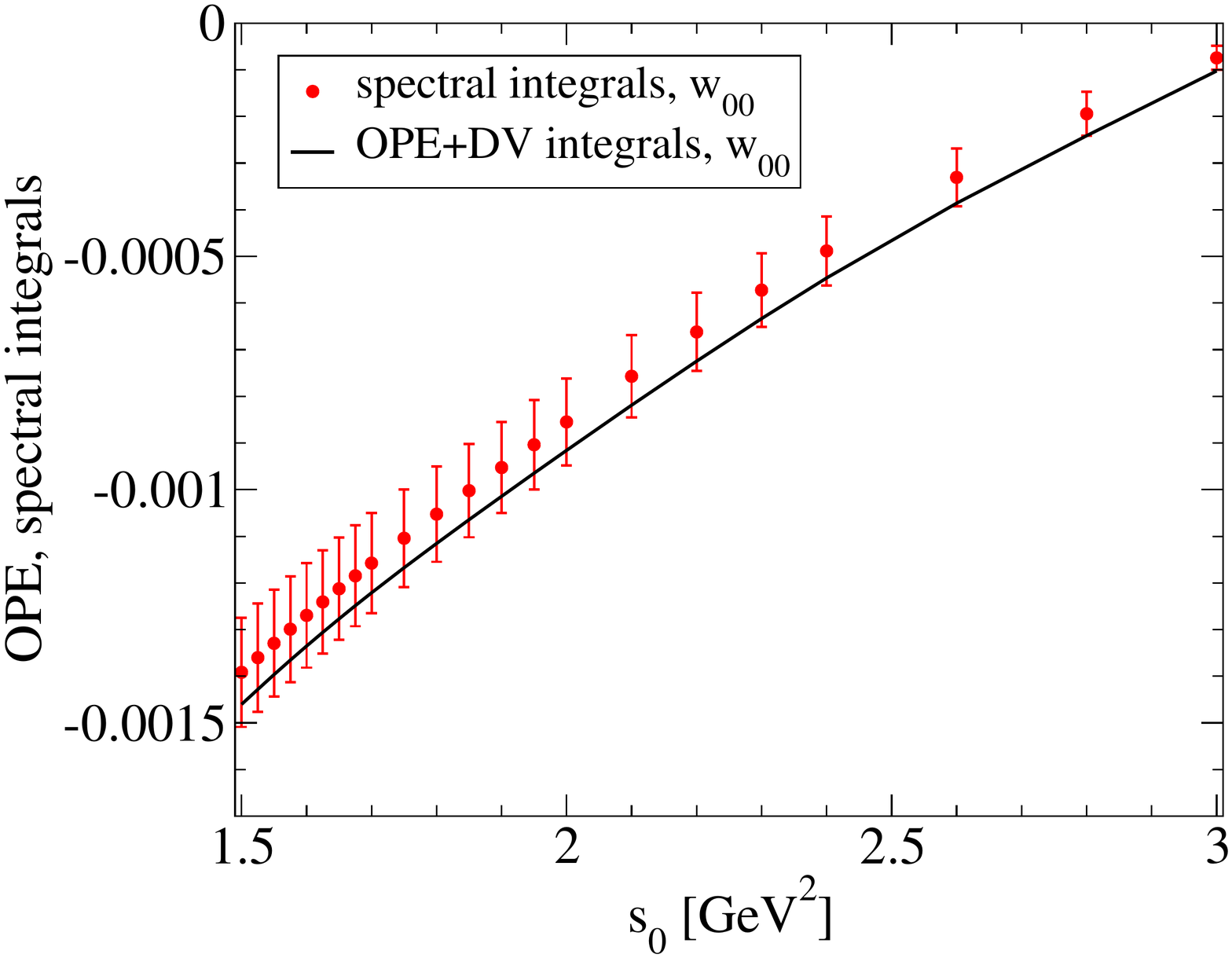}
\hspace{-0.9cm}
\includegraphics*[width=7.9cm]{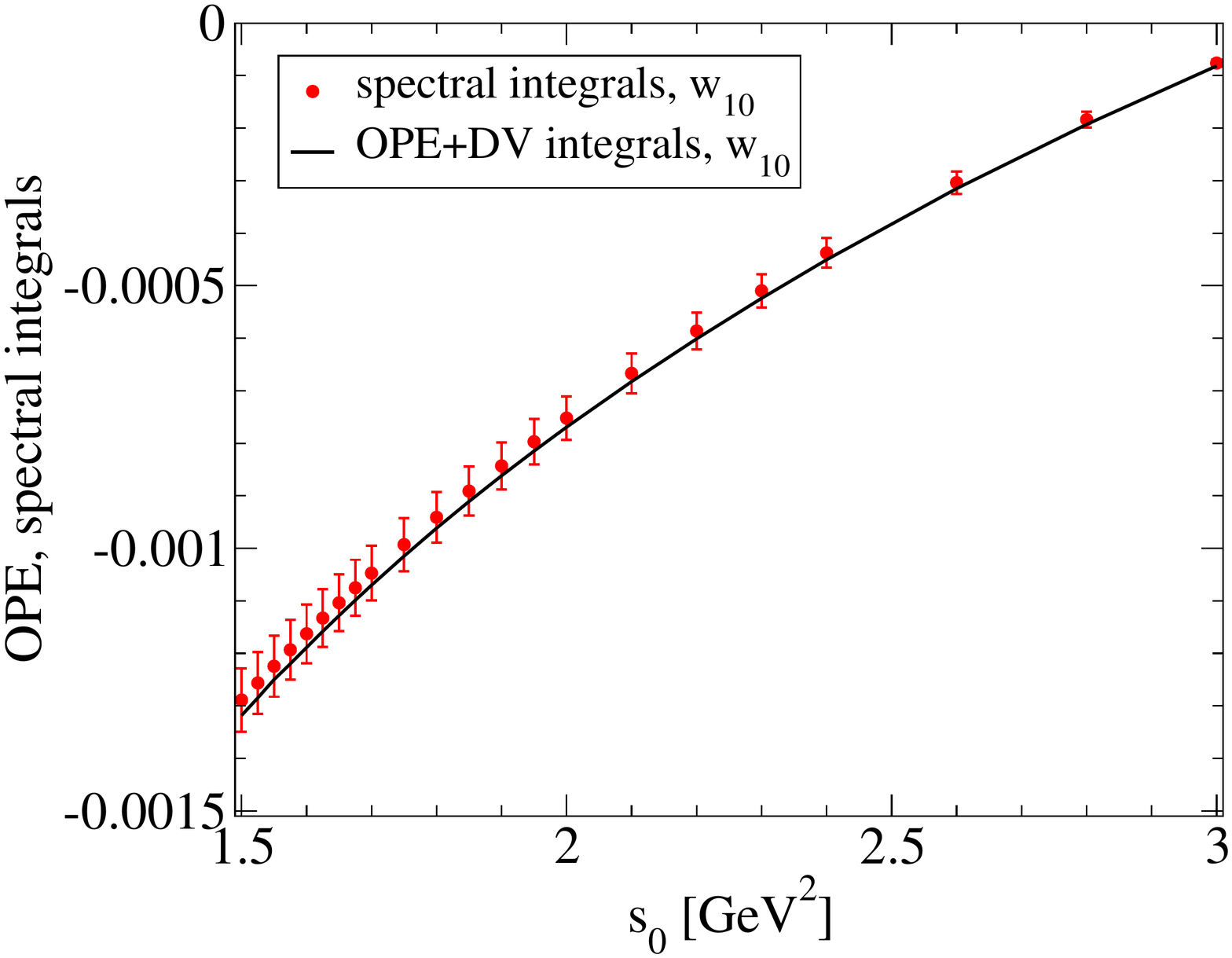}
\vspace{0.0cm}
\includegraphics*[width=7.9cm]{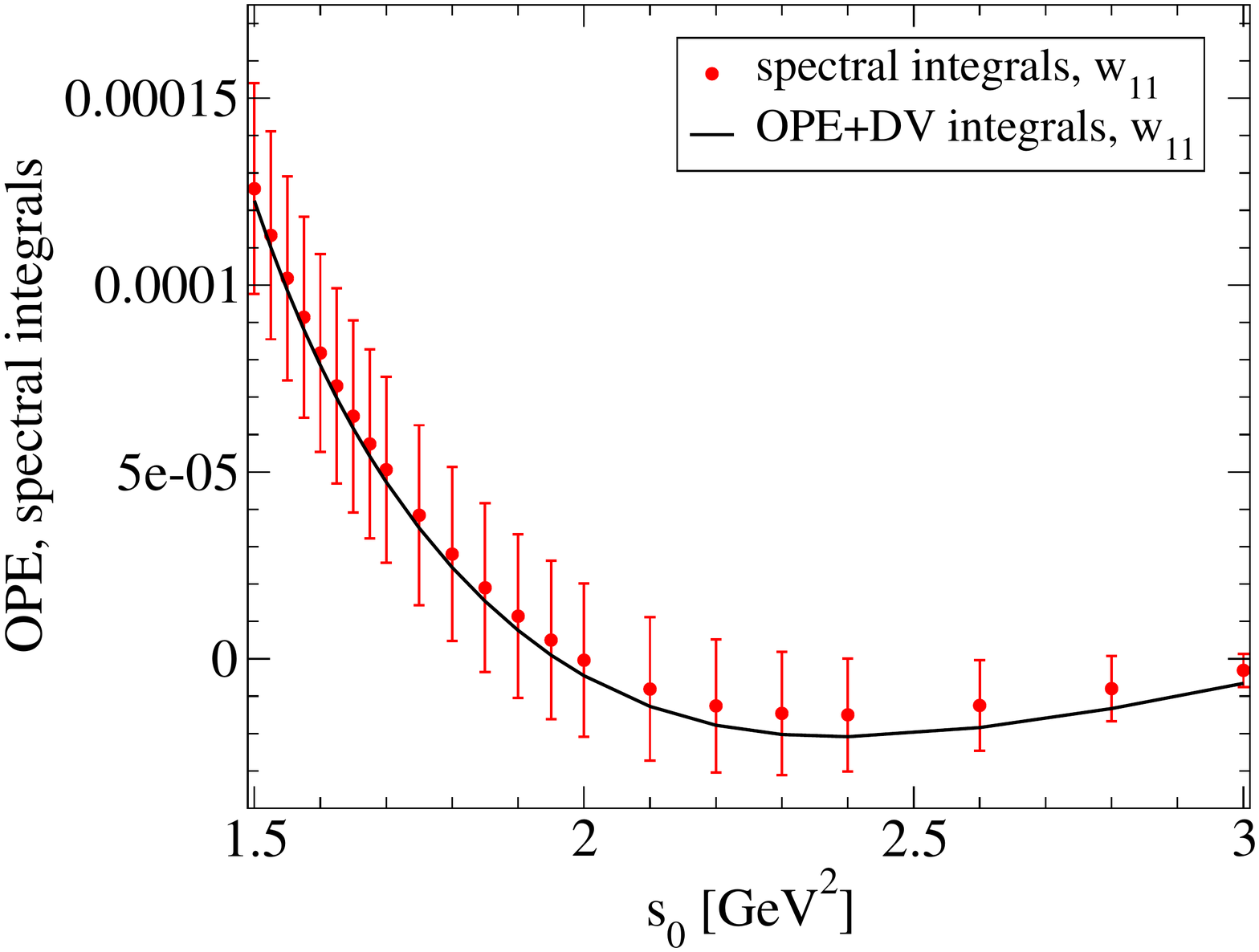}
\hspace{-0.9cm}
\includegraphics*[width=7.9cm]{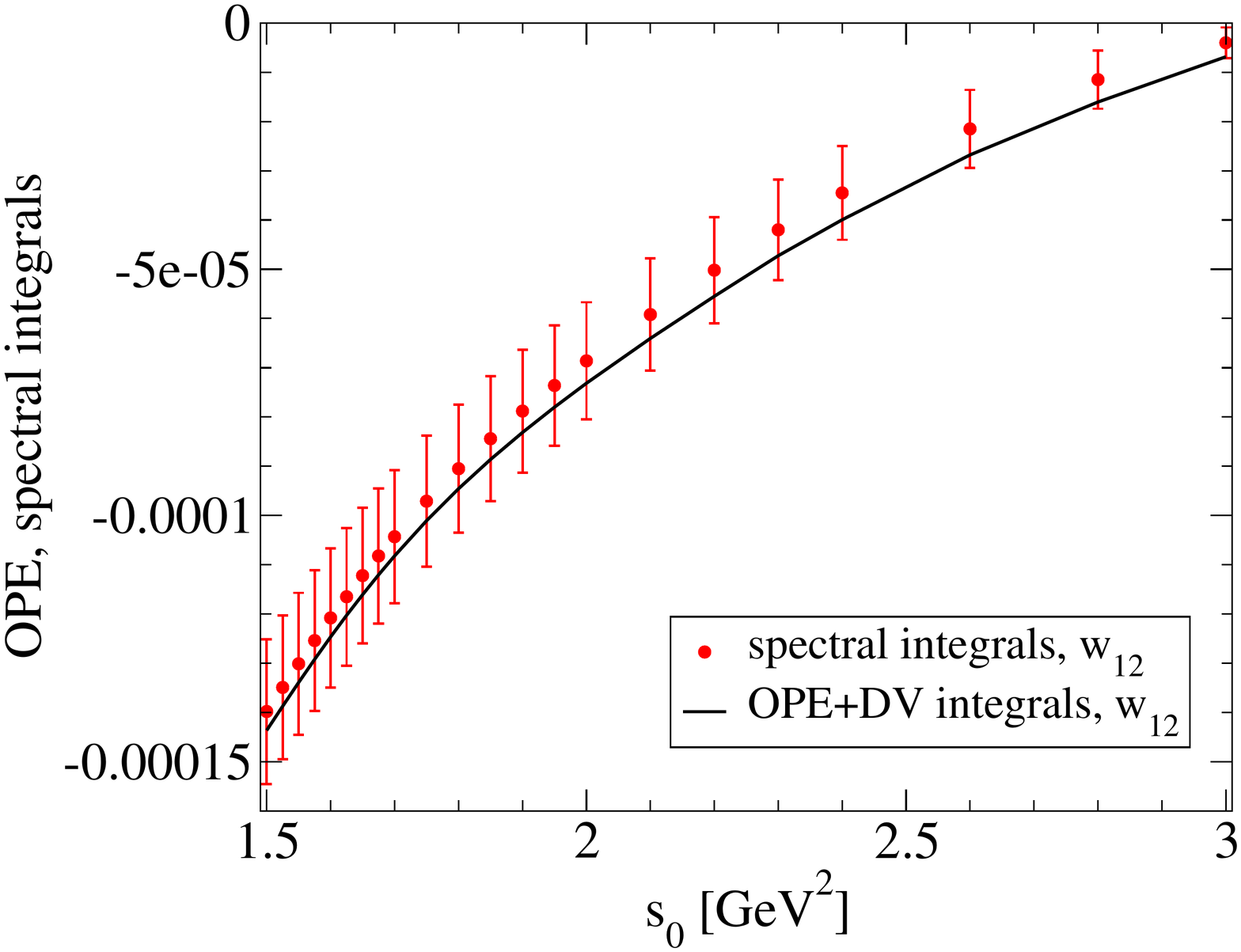}
\vspace{0.0cm}
\includegraphics*[width=7.9cm]{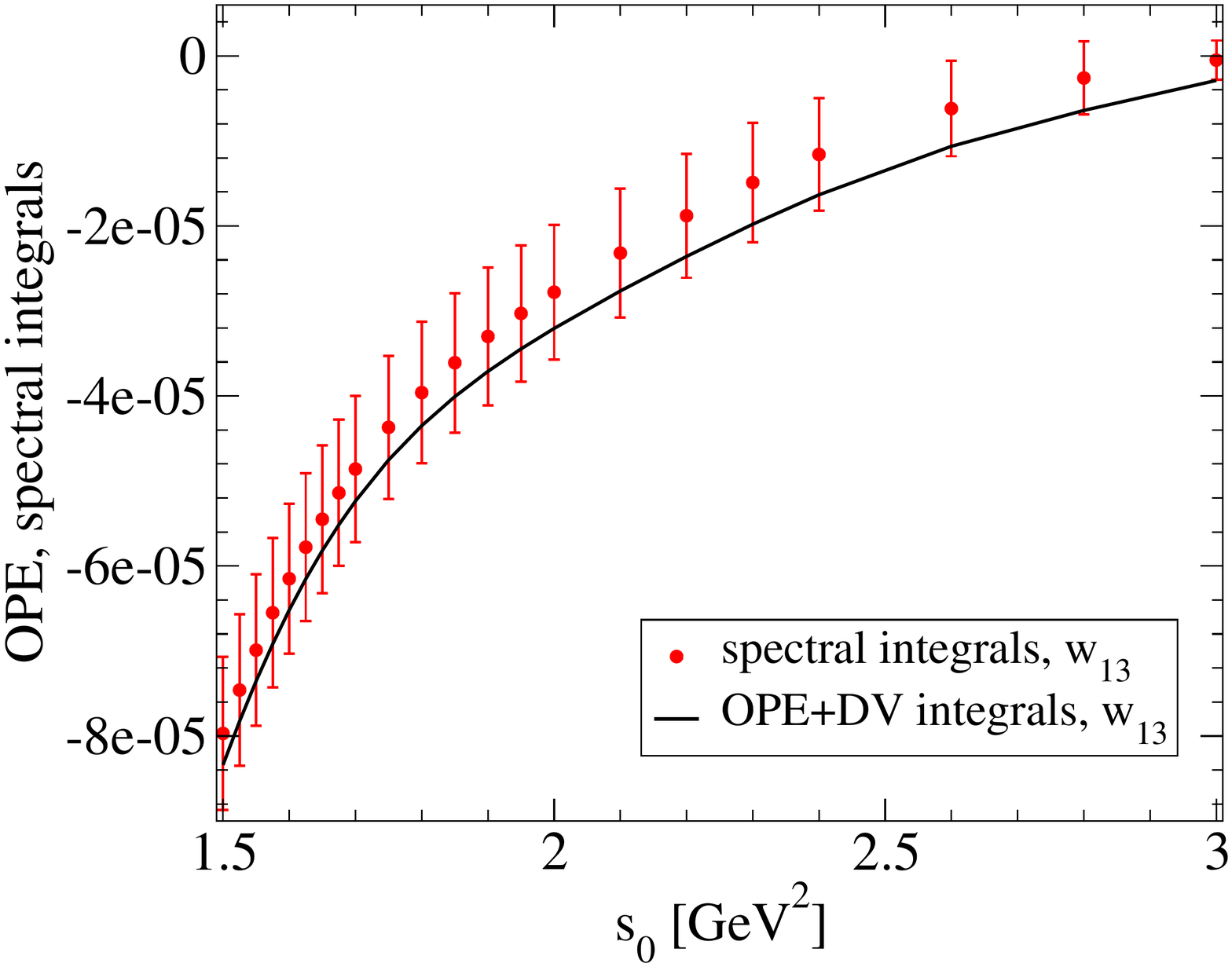}
\end{center}
\begin{quotation}
\vspace*{-4ex}
\floatcaption{wklus}%
{\it Comparison of $V+A$ spectral integrals
with weights $w_{k\ell}$ of Eq.~(\ref{ALEPH}), using results of the CIPT fit with
$s_{\rm min}=1.55$~{\rm GeV}$^2$ of Table~5 of Ref.~\cite{alphas14}
with data, using the ALEPH data of Ref.~\cite{ALEPH13}.}
\end{quotation}
\vspace*{-4ex}
\end{figure}
We note that, of the moments shown in Fig.~\ref{wklus},
only $w_{00}$ was used explicitly in our fit. The OPE coefficients $C_D$, 
$D>8$, required to obtain the theoretical moments for the weights 
$w_{10}$, $w_{11}$, $w_{12}$ and $w_{13}$, were computed using the power 
weight $x^N$ FESRs at a single $s_0$ (chosen equal to $1.55$~GeV$^2$)
with our $\alpha_s$ and DV parameter
fit results as input~\cite{alphas14}. The resulting $C_D$ values are listed 
in Eq.~(\ref{OPE74}). The agreement of the $w_{10}$, $w_{11}$, $w_{12}$ and 
$w_{13}$ spectral integrals with the corresponding theoretical 
representations, as a function of $s_0$, thus provides a test of 
the self-consistency of our strategy.

Figure~\ref{wklPR} shows the same type of plots, but now using the results 
from the CIPT fit given in Table~\ref{tab1} (which corresponds to Table~1 of 
Ref.~\cite{Pich}), obtained ignoring DVs, and setting 
$C_{10}=C_{12}=C_{14}=C_{16}=0$.
\begin{figure}[t!]
\begin{center}
\includegraphics*[width=7.9cm]{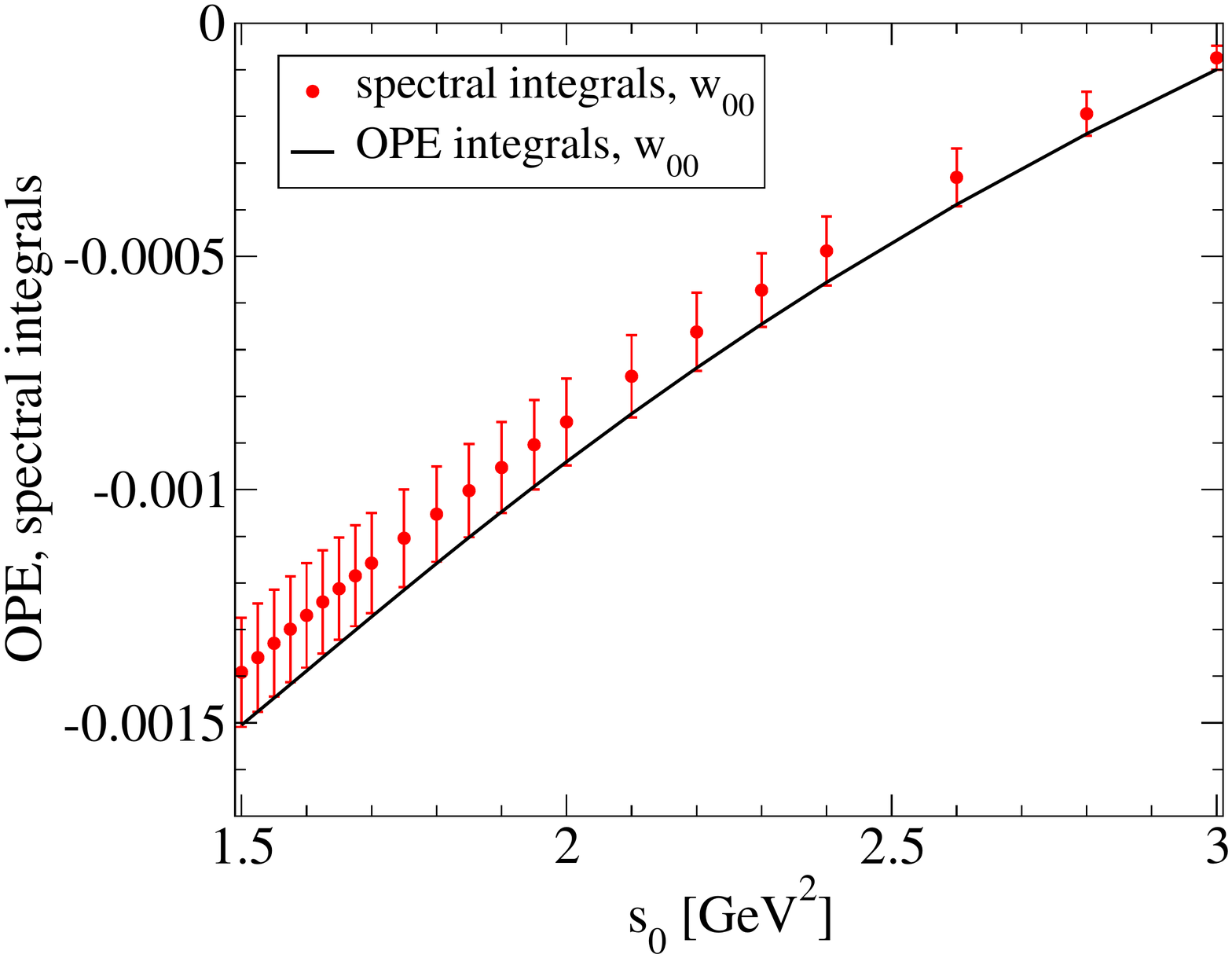}
\hspace{-0.9cm}
\includegraphics*[width=7.9cm]{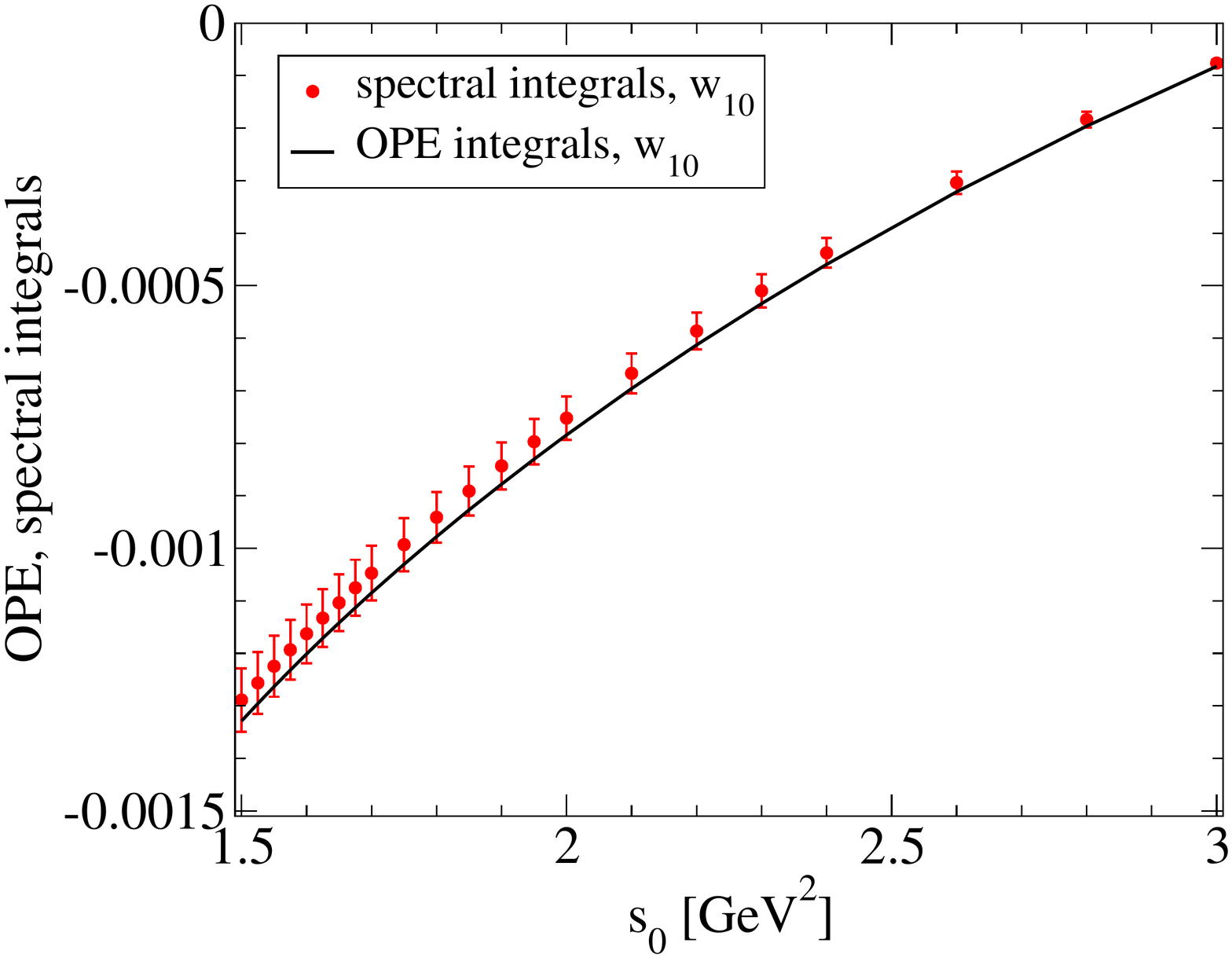}
\vspace{0.0cm}
\includegraphics*[width=7.9cm]{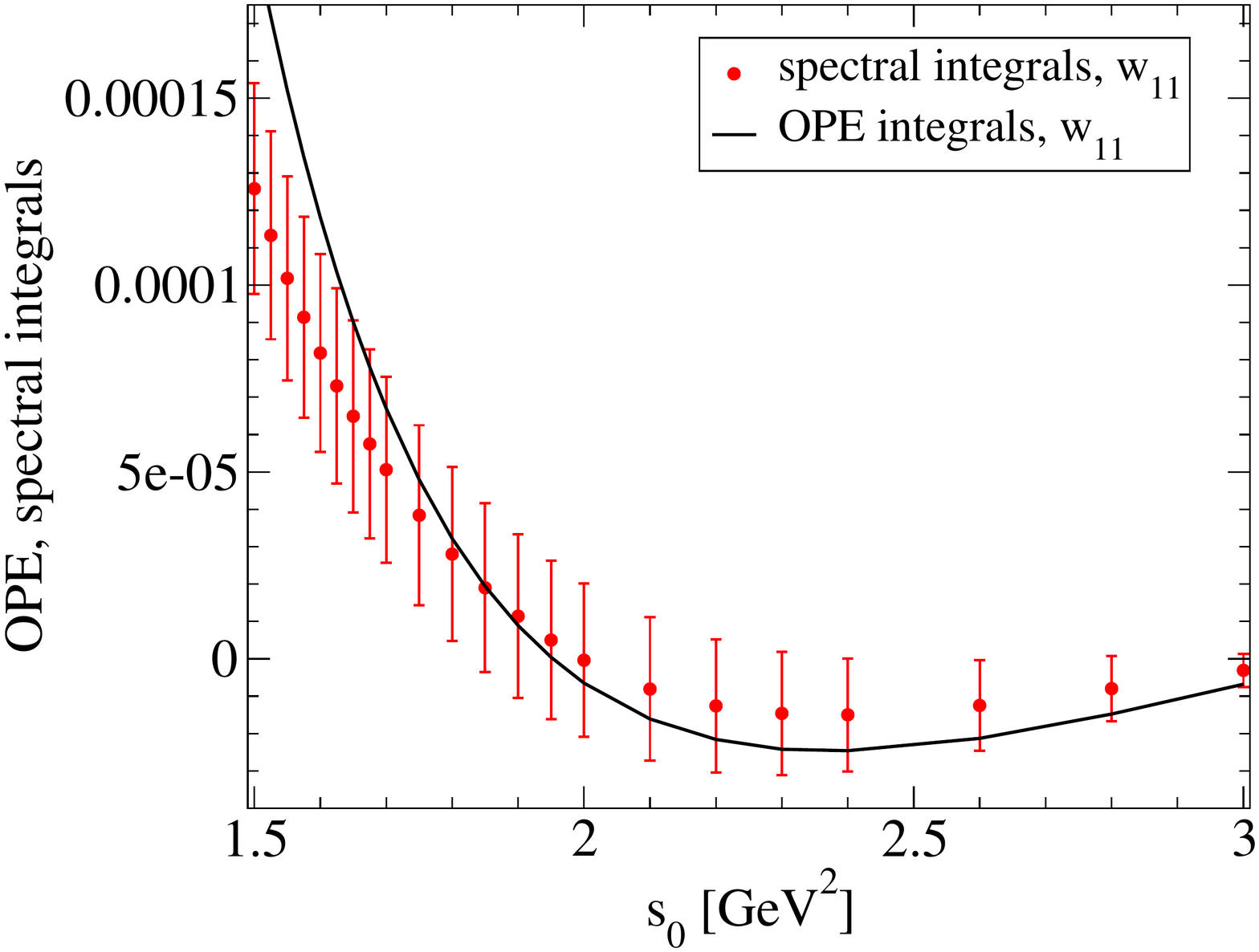}
\hspace{-0.9cm}
\includegraphics*[width=7.9cm]{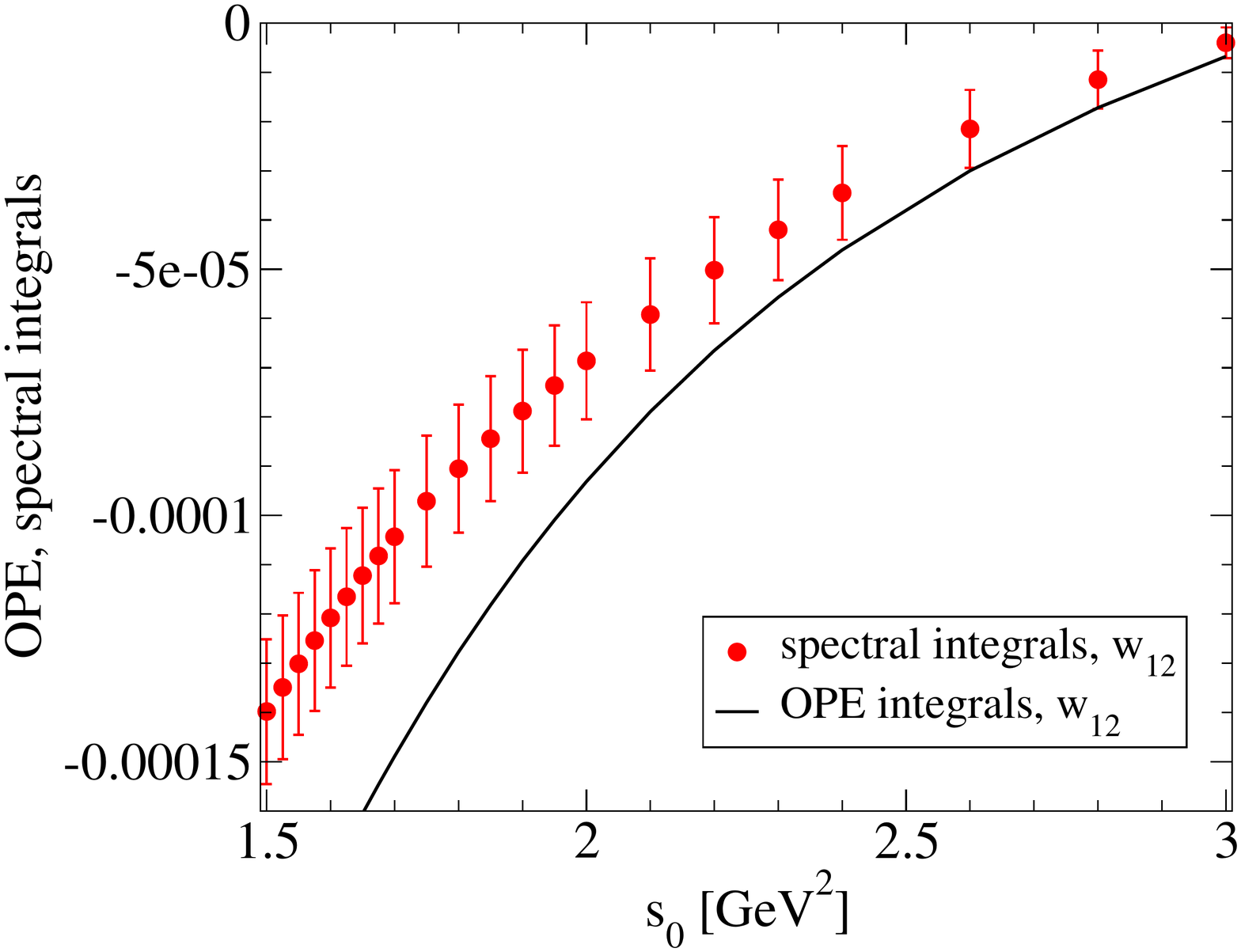}
\vspace{0.0cm}
\includegraphics*[width=7.9cm]{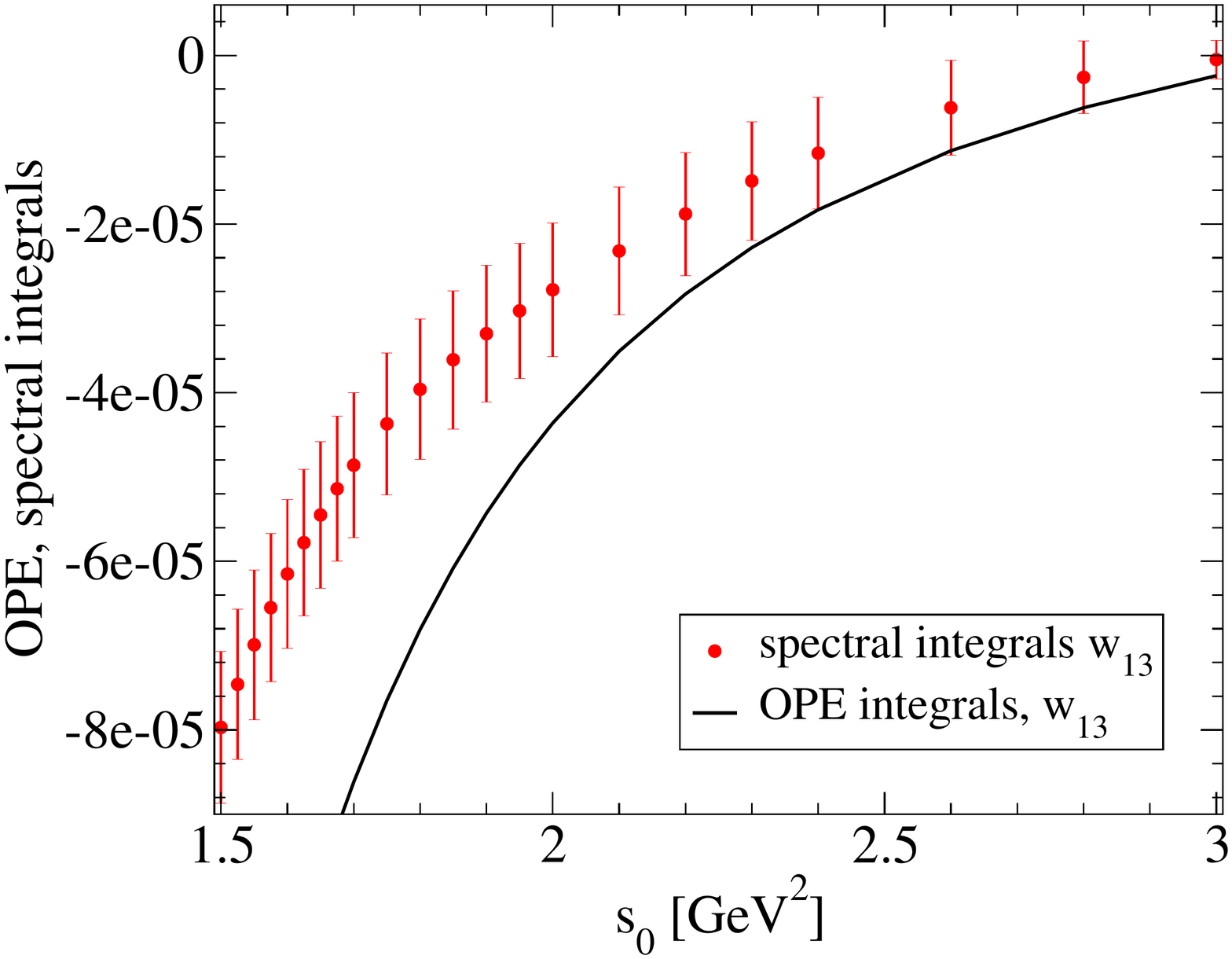}
\end{center}
\begin{quotation}
\vspace*{-4ex}
\floatcaption{wklPR}%
{\it Comparison of $V+A$ spectral integrals
with weights $w_{k\ell}$ of Eq.~(\ref{ALEPH}), using results of the CIPT fit in
Table~\ref{tab1}, \ie, Table~1 of Ref.~\cite{Pich}
with data, using the ALEPH data of Ref.~\cite{ALEPH13}.}
\end{quotation}
\vspace*{-4ex}
\end{figure}
Note that in this case all spectral integrals at $s_0=m_\t^2$ were included
in the fit. One clearly sees that the $s_0$ dependence deteriorates for weights
which probe the higher-dimension terms in the OPE. The comparison
of Figs.~\ref{wklus} and \ref{wklPR} clearly favors the DV-model strategy over
the truncated-OPE strategy.

We have also considered examples of FOPT fits, again using our fit for
$s_{\rm min}=1.55$~GeV$^2$ from Table~5 of Ref.~\cite{alphas14} and the fit of
Table~1 of Ref.~\cite{Pich}.\footnote{For these plots, we estimated FOPT values
for $C_{10-16}$ analogous to the CIPT estimates given in Ref.~\cite{alphas14}.} We show the spectral integrals with weights
$w_{10}$ and $w_{13}$ in Fig.~\ref{FOPT}. The two weights we chose
to show are representative of the whole set, except for $w_{00}$ for which
the DV-model-strategy plot looks as good as in the CIPT case. This is no
surprise, as the $s_0$ dependence of the spectral integral with weight $w_{00}$
was used in the fits based on this strategy.
\begin{figure}[t!]
\begin{center}
\includegraphics*[width=7.9cm]{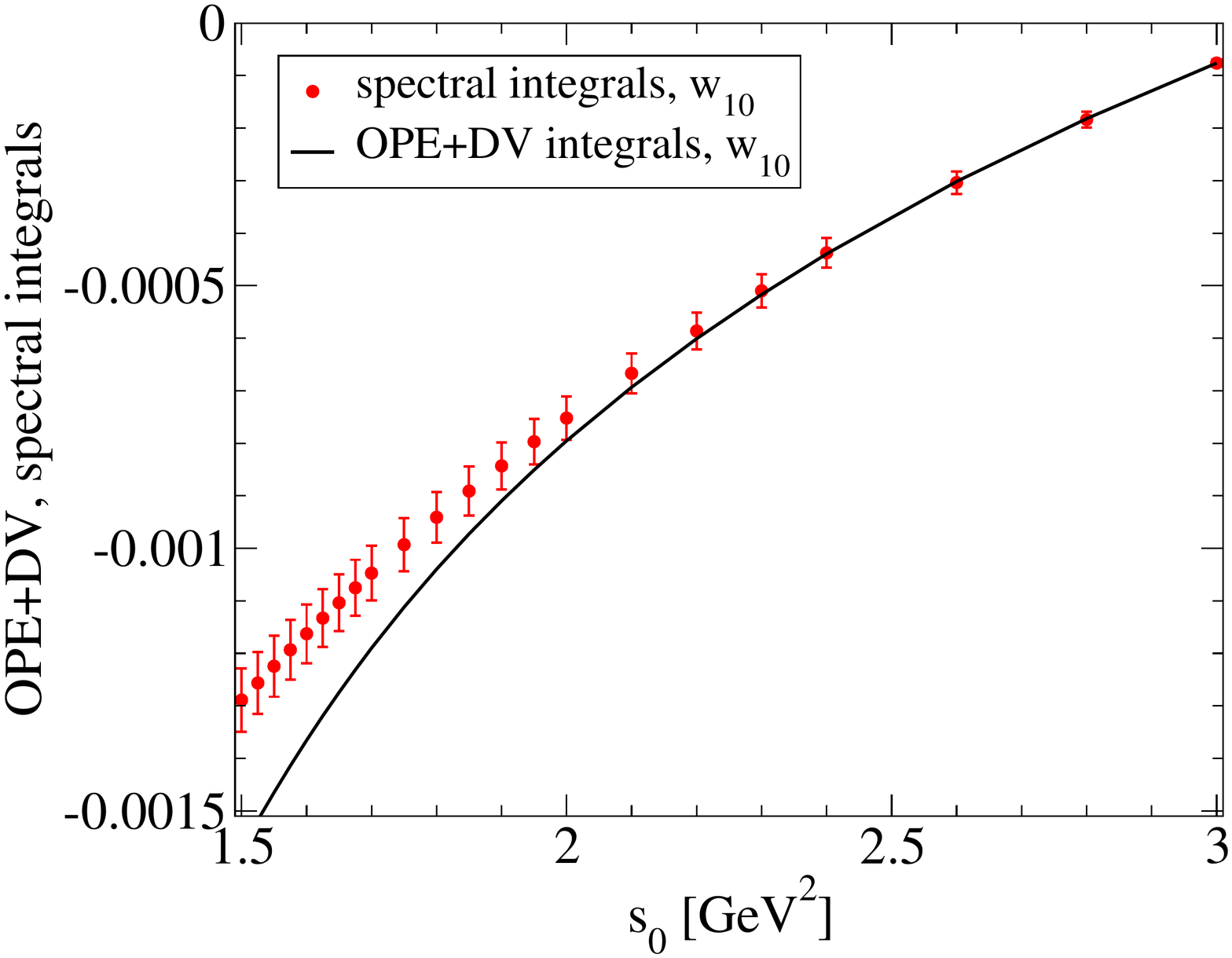}
\hspace{-0.9cm}
\includegraphics*[width=7.9cm]{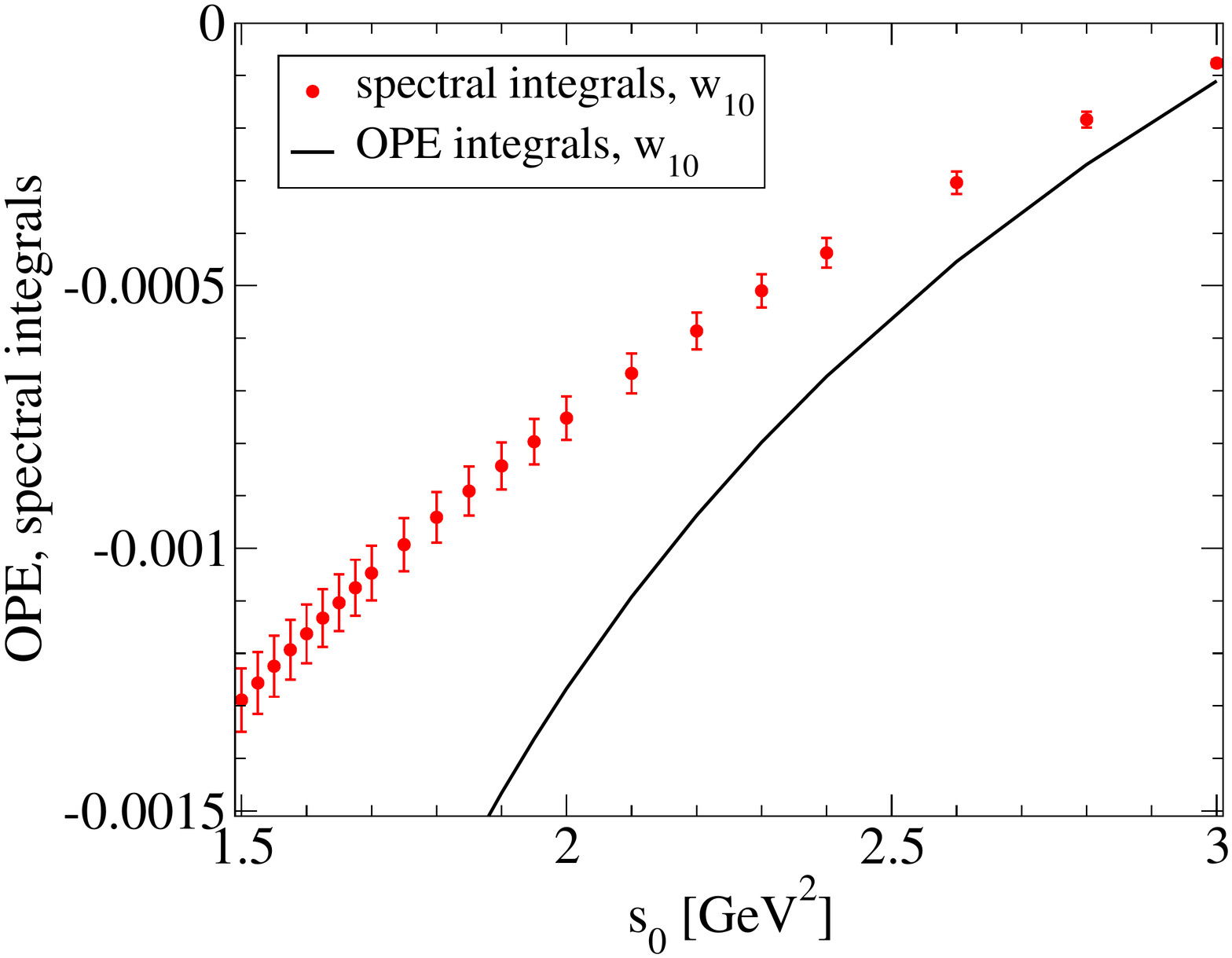}
\vspace{0.0cm}
\includegraphics*[width=7.9cm]{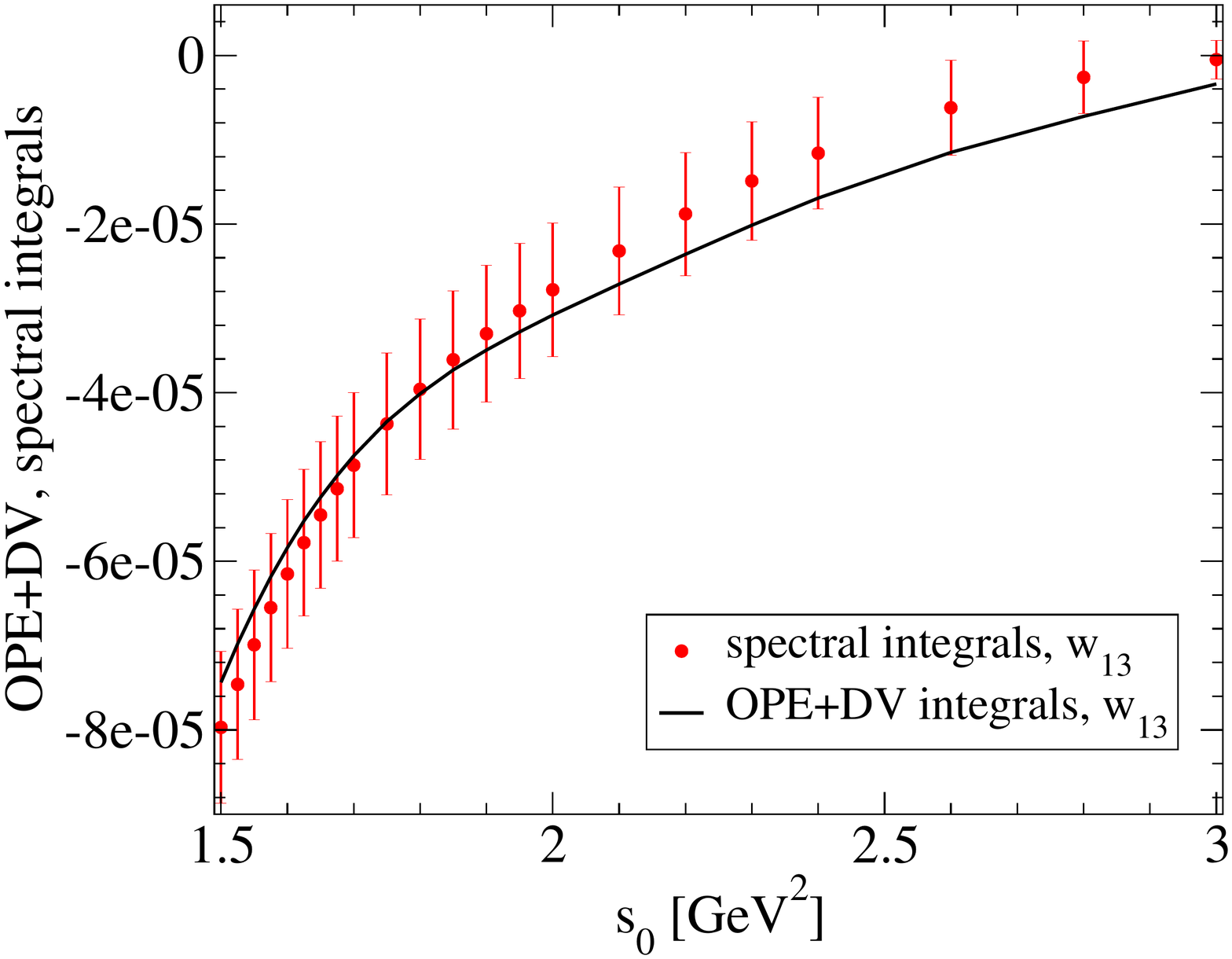}
\hspace{-0.9cm}
\includegraphics*[width=7.9cm]{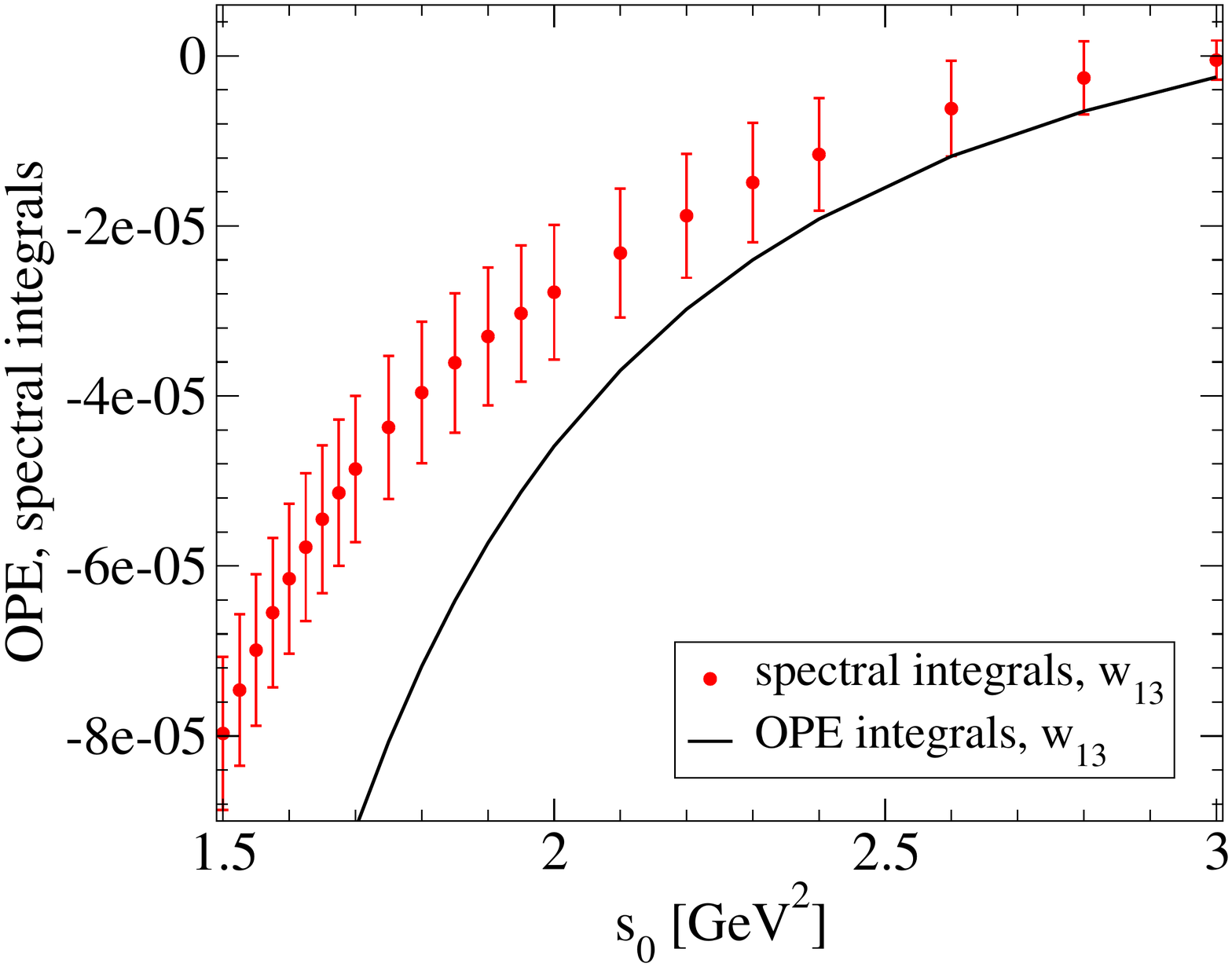}
\end{center}
\begin{quotation}
\vspace*{-4ex}
\floatcaption{FOPT}%
{\it Comparison of $V+A$ spectral integrals
with weights $w_{k\ell}$ of Eq.~(\ref{ALEPH}), using results of the FOPT fit with
$s_{\rm min}=1.55$~{\rm GeV}$^2$ of Table~5 of Ref.~\cite{alphas14} (left panels),
and using results of the FOPT fit in
Table~\ref{tab1}, \ie, Table~1 of Ref.~\cite{Pich} (right panels),
with data, using the ALEPH data of Ref.~\cite{ALEPH13}.}
\end{quotation}
\vspace*{-4ex}
\end{figure}
Although the performance of the DV model in the FOPT case is somewhat worse 
than in the CIPT case, it is still much better than that of
the truncated-OPE model.  We note that the results for
$\a_s(m_\t^2)$ obtained with the DV-model strategy in Ref.~\cite{alphas14}
do not rely on spectral integrals with weights $w_{1\ell}$, whereas
all these weights are used in the truncated-OPE strategy.

\subsection{\label{criticism} The criticism of Ref.~\cite{Pich}}
In the preceding sections, we have demonstrated that the truncated-OPE-model
strategy suffers from systematic problems which preclude its use as
a method for obtaining a reliable determination of $\a_s(m_\t^2)$ from 
hadronic $\t$ decays. It is, however, also relevant to ask whether the 
DV-model strategy provides an acceptable alternative, and Ref.~\cite{Pich} 
devoted a section to criticism of this strategy. The key criticisms raised by 
Ref.~\cite{Pich} are encapsulated in Fig.~6 and Table~10 of Ref.~\cite{Pich}. 
Here we will address these criticisms, both refuting them and at the
same time commenting more specifically on some of their more 
misleading aspects.

First, we consider the argument based on
Fig.~6 of Ref.~\cite{Pich}, which chooses to focus on the simplest,
but non-central, fit of Ref.~\cite{alphas14}.
We reproduce this figure in Fig.~\ref{DVmodelV}. The 
 fit considered in Ref.~\cite{Pich} is a fit of 
perturbation theory (FOPT) and the DV {\it ansatz}~(\ref{ansatz}) to the
$s_0$-dependent ($w=1$)-weighted integrals of the $V$ spectral 
function on the left-hand side of Eq.~(\ref{sumrule}).\footnote{According to 
Eq.~(\ref{degreecond}) no OPE coefficients $C_{D\ge 2}$ are probed for
the choice $w=1$.}  The fit is performed in the interval
$[s_{\rm min},m_\t^2]$. Figure~\ref{DVmodelV} shows the resulting
$\a_s(m_\t^2)$ (left panel) and the $p$-value of the fit (right panel), 
as a function of $s_{\rm min}$.  This figure  is in good agreement with
Fig.~6 of Ref.~\cite{Pich}. There are small differences; in particular, 
our $p$-values are somewhat higher, likely as a result of the somewhat 
more careful treatment of the ALEPH data in Ref.~\cite{alphas14}
(\seef\ footnote~6).

Let us now explain why the criticism of Ref.~\cite{Pich} based on 
these two plots is unjustified. With regard to the left panel of 
Fig.~\ref{DVmodelV}, Ref.~\cite{Pich} states that ``the fitted values of 
$\a_s(m_\t^2)$ do not present the stability one would expect.'' This is 
a misreading of the plot. By varying $s_{\rm min}$, as was also done in 
Ref.~\cite{alphas14}, one lets the data decide whether a stability region 
exists. To the left of this region (if it exists), \ie, for smaller $s_{\rm min}$,
the importance of non-perturbative effects not captured by perturbation theory
or the DV {\it ansatz}~(\ref{ansatz}) causes the value of $\a_s(m_\t^2)$ to 
move. For larger values of $s_{\rm min}$, there should be stability, but, as 
fewer points are included in the fit if $s_{\rm min}$ increases, the results 
will become noisier. This is precisely what one sees in Fig.~\ref{DVmodelV}, 
and there is, moreover, a nice ``plateau'' (region
of stability) for $1.5$~GeV$^2\,\ltap\, s_{\rm min}\,\ltap\, 1.8$~GeV$^2$.

\begin{figure}[t!]
\begin{center}
\includegraphics*[width=7.4cm]{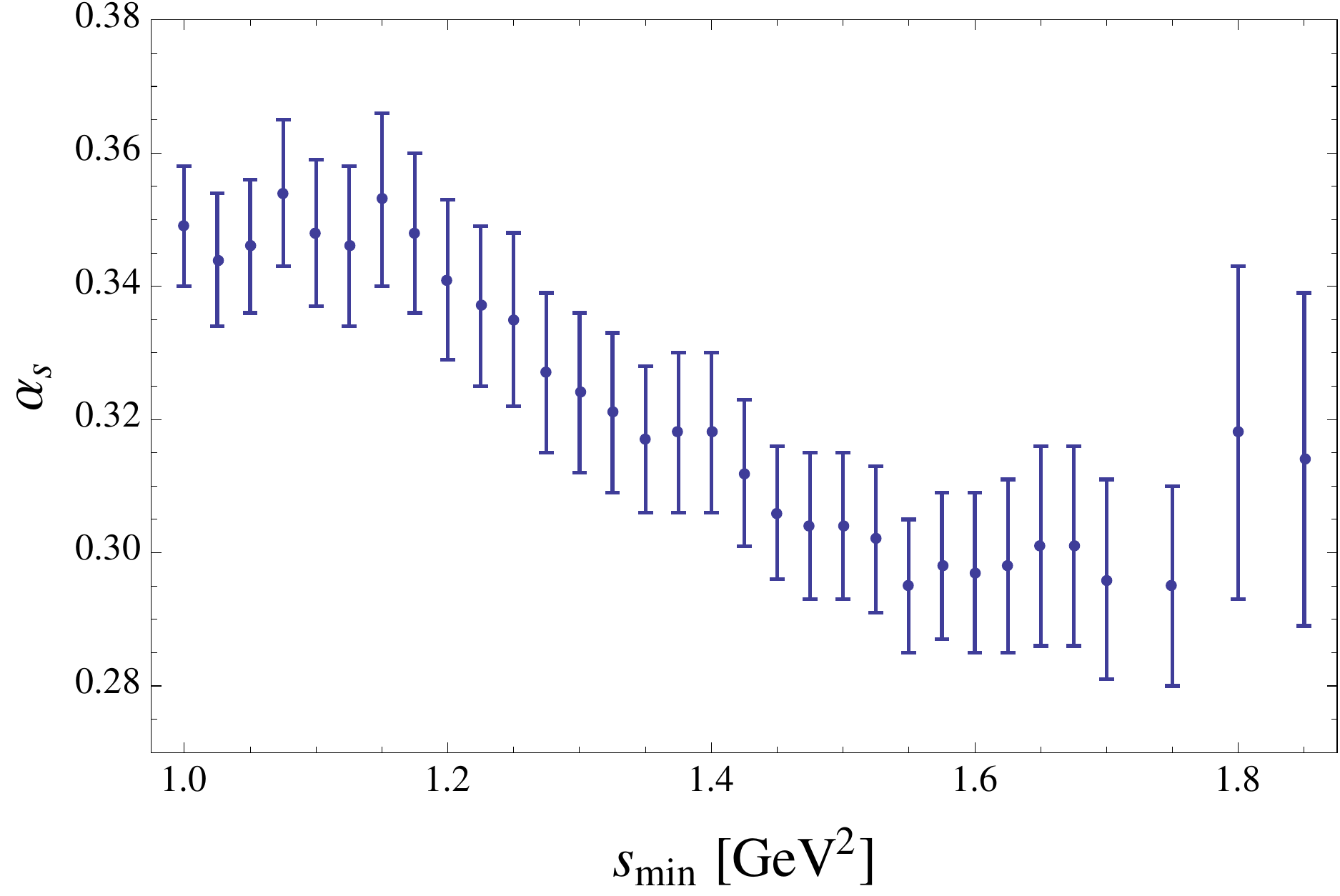}
\hspace{0.0cm}
\includegraphics*[width=7.4cm]{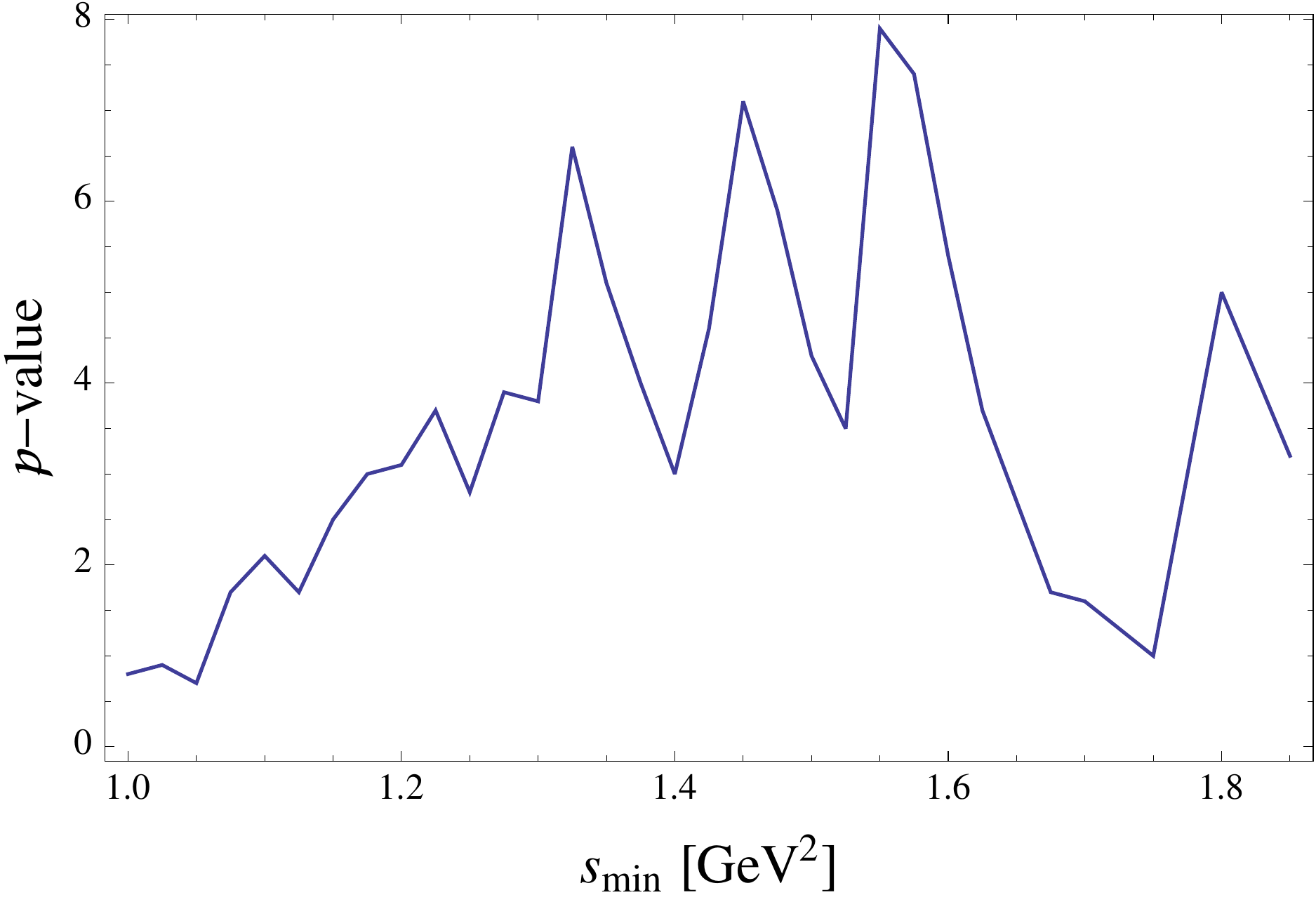}
\end{center}
\begin{quotation}
\vspace*{-4ex}
\floatcaption{DVmodelV}%
{\it FOPT determination of $\a_s(m_\t^2)$ as a function of $s_{\rm min}$, using 
the $V$-channel $w=1$ fit of Ref.~\cite{alphas14}.}
\end{quotation}
\vspace*{-4ex}
\end{figure}
With regard to the $p$-values shown in the right-hand panel, Ref.~\cite{Pich} 
states that ``If the model were reliable, it should work better at higher 
hadronic invariant masses,'' and takes both the size of the $p$-values 
in the region of the plateau, and the ``significant deviations'' at 
larger $s_{\rm min}$ as a signal of ``poor statistical quality.'' 
One should bear in mind, however, that the $p$-value of a fit is {\it itself}
a statistical quantity, that will fluctuate with the data. For larger 
$s_{\rm min}$, the fluctuations in the data are more pronounced, and the 
$p$-values follow suit. Furthermore, a $p$-value of about 8\% is 
generally not regarded as a proof of the failure of a hypothesis 
in a statistical analysis. In Ref.~\cite{alphas14} many other fits where 
carried out (including multiple-weight V channel and
combined V and A channel fits, and a combined $w=1$ $V$ and $A$ channel
fit with $p$-values about double those of the corresponding $V$-channel-only 
fit. The results were also subjected to further tests, such as those provided
by the Weinberg sum rules). The internal consistency of all these 
tests was taken as evidence for the likely validity of the final result.

\begin{figure}[t!]
\begin{center}
\includegraphics*[width=7.4cm]{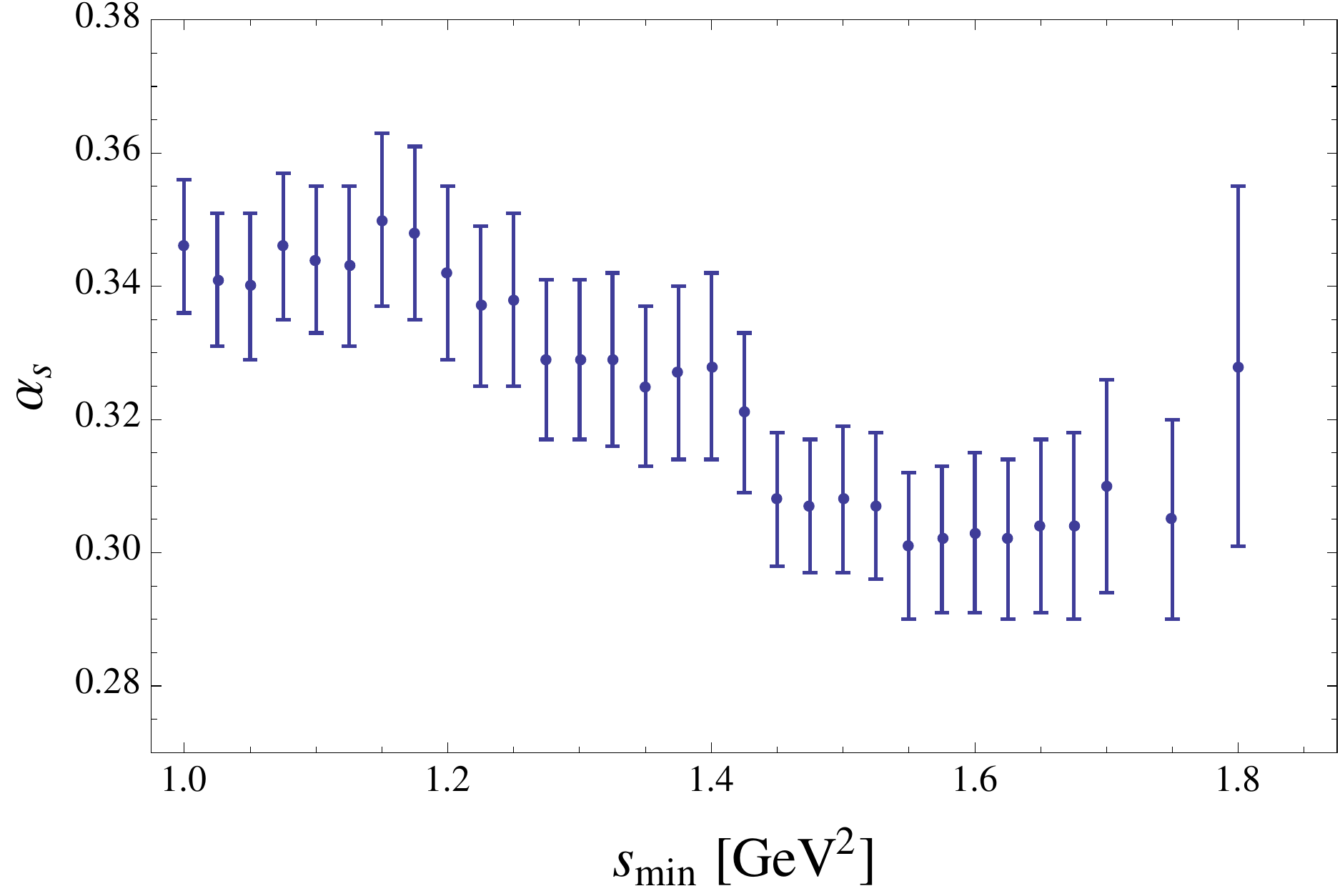}
\hspace{0.0cm}
\includegraphics*[width=7.4cm]{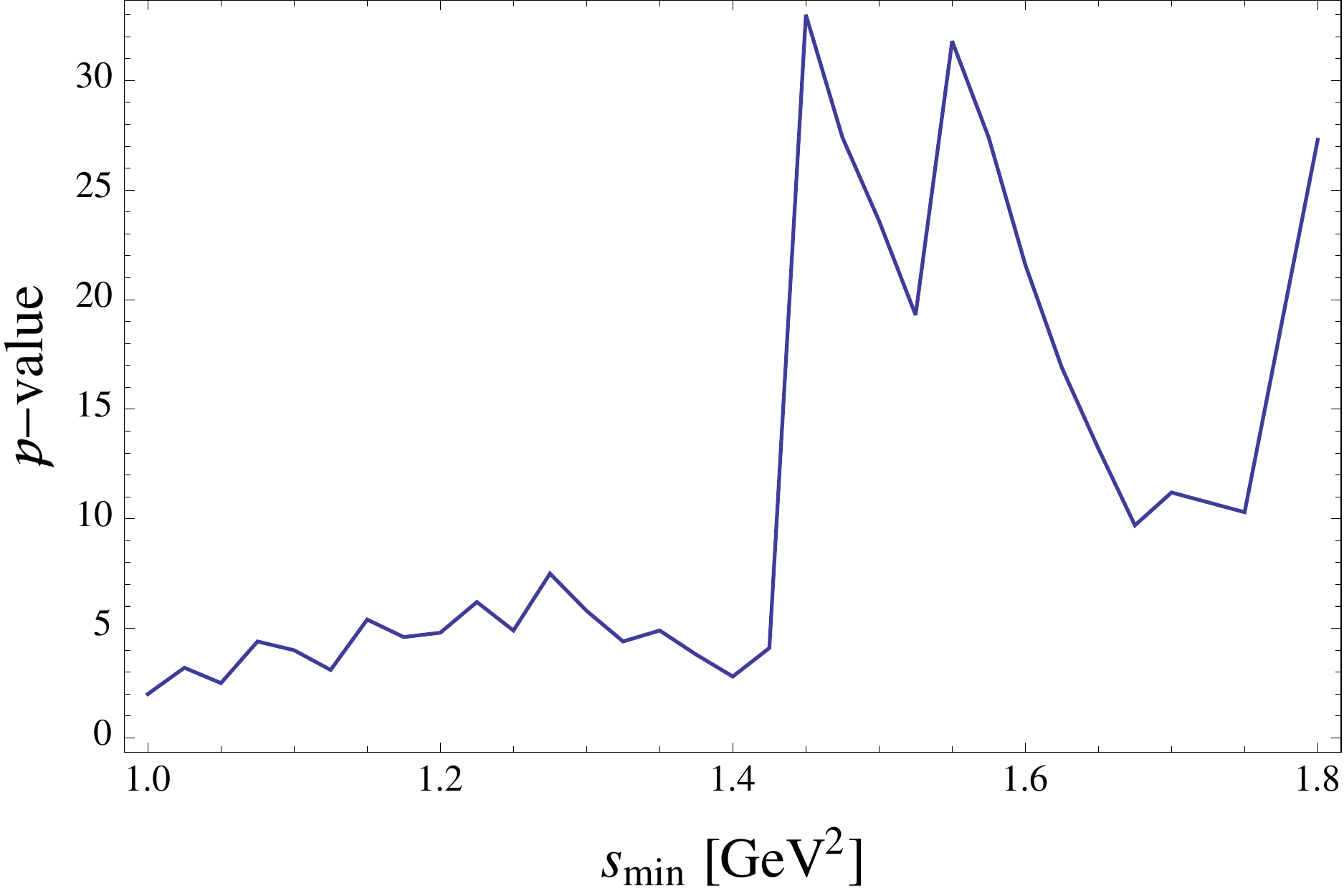}
\end{center}
\begin{quotation}
\vspace*{-4ex}
\floatcaption{DVmodelVfake}%
{\it FOPT determination of $\a_s(m_\t^2)$ as a function of $s_{\rm min}$, using the $V$-channel $w=1$ fit of Ref.~\cite{alphas14},
but now replacing the ALEPH data with fake data (see text).}
\end{quotation}
\vspace*{-4ex}
\end{figure}

We may further test this reasoning by repeating the DV-model analysis 
on a fake data set, as in Sec.~\ref{fake data}.  To do so, we generated 
a fake data set in the same way as in Sec.~\ref{generation}, but now for the 
$V$ channel only, using, to be specific, FOPT perturbation theory. 
The results of this exercise are shown in Fig.~\ref{DVmodelVfake}.
The patterns are the same as those seen in Fig.~\ref{DVmodelV}. 
The only difference is that, in the case of Fig.~\ref{DVmodelVfake}, 
we know that the fake data has been constructed from a theoretical spectral 
function with $\a_s(m_\t^2)=0.297$ and the DV parameters given in the 
$s_{\rm min}=1.55$~GeV$^2$ FOPT line of Table~IV in Ref.~\cite{alphas14}.

Considering first the left panel of Fig.~\ref{DVmodelVfake}, we see 
that the pattern is consistent with the pattern seen in the 
left panel of Fig.~\ref{DVmodelV}. In addition, the
fits find the correct values of $\a_s(m_\t^2)$ and the DV parameters
$\d_V$, $\g_V$, $\a_V$ and $\b_V$ within errors.\footnote{At 
$s_{\rm min}=1.55$~GeV$^2$, the fit finds the value $0.301(11)$, where 
the error is statistical only.} The stability plateau is located in the 
same $s_{\rm min}$ range in both figures. Also the right panels
look very similar. For $s_{\rm min}\,\ltap\, 1.4$~GeV$^2$, the fits to
both the real and fake data have similar $p$-values, indicating that the 
theory representation works less well in that region, and stops working 
toward lower $s_{\rm min}$. For larger $s_{\rm min}$, the $p$-values of the 
fake-data fits are better than for the real-data fits. This is no
surprise: after all, the theory is ``perfect'' for the fake data, 
whereas it is not expected to be so for the real data. More 
importantly, the large downward fluctuation near $s_{\rm min}=1.7$~GeV$^2$
is seen in both figures. By construction of the fake-data fits, the 
putative conclusion that these would be bad fits for the fake-data 
case is obviously incorrect. Therefore, the same conclusion 
cannot be drawn for the real-data fits either: it is not excluded 
that the feature seen around $s_{\rm min}=1.7$~GeV$^2$ in the right panel 
of Fig.~\ref{DVmodelV} is nothing else than a fluctuation.

Turning now to the second point, let us briefly comment on Table~10 
of Ref.~\cite{Pich}. Again, Ref.~\cite{Pich} considers the non-central, FOPT fit of the 
spectral integral with weight $w(x)=1$ of Ref.~\cite{alphas14}. However, 
instead of using the DV {\it ansatz} as given in Eq.~(\ref{ansatz}), it is now
multiplied by $s^n$, with $n=0,\ 1,\ 2,\ 4,\ 8$ ($n=0$ corresponds to 
Eq.~(\ref{ansatz}), of course). This makes very little sense, as follows from 
the discussion of the {\it ansatz}~(\ref{ansatz}) at the end of 
Sec.~\ref{theory}. While Eq.~(\ref{ansatz}) introduces a particular model 
for DVs (which are a manifestation of the resonances one sees in the 
spectral functions), one should insist that any model incorporates what we
know about the phenomenon the {\it ansatz} is supposed to model. In 
particular, Sec.~\ref{theory} suggests that one might try to introduce 
a prefactor $1+a_1/s+a_2/s^2+\dots$, with new parameters $a_{1,2,\dots}$,
reflecting the expectation that Eq.~(\ref{ansatz}) emerges because of the suspected
asymptotic nature of the OPE, for which the expansion parameter is $1/s$. The
multiplication of Eq.~(\ref{ansatz}) with an {\it inverse} power of this expansion
parameter is, in view of these expectations, a wildly arbitrary choice, with 
no root in anything we know or suspect about the physics of QCD.

Despite the obvious shortcomings of the modifications to the DV ansatz 
employed in Ref.~\cite{Pich}, the fits shown in Table~10 of 
Ref.~\cite{Pich}, in fact, show a remarkable {\it stability} as a 
function of $n$. The $p$-values shown there are essentially constant. 
In terms of the criteria of Ref.~\cite{Pich}, where $1.5$ $\sigma$ 
central-value shifts are deemed indications of stability of the analysis, 
one would have to characterize the results for $\a_s(m_\t^2)$ as 
surprisingly constant: even the value at $n=8$, $\a_s(m_\t^2)=0.314(15)$, 
is just 1~$\s$ away from the central $n=0$ value $0.298$, while that at 
$n=4$ is closer to $0.5\sigma$ distant. The conclusion that such
variations prove ``model dependence,'' as claimed in Ref.~\cite{Pich}, thus 
seems to us a somewhat surprising one.  
While we believe that the attempt of Ref.~\cite{Pich} to vary the DV model 
is theoretically unfounded and thus quite arbitrary, the tests of 
Table~10 in Ref.~\cite{Pich} in fact only serve to confirm the 
reliability of the results of Ref.~\cite{alphas14}.

\section{\label{conclusion} Conclusion}
Our main goal in this article was a thorough investigation of the 
truncated-OPE-model strategy for extracting $\a_s(m_\t^2)$ from
experimental data for hadronic $\t$ decays. This strategy has been used 
extensively, notably in Refs.~\cite{ALEPH13,Pich,ALEPH,OPAL}. In a series of 
articles \cite{alphas1,alphas2,alphas14} we have designed and implemented 
a different strategy, the DV-model strategy. This strategy gives 
different results for $\a_s(m_\t^2)$, and it is thus important to 
understand and appraise the difference. In Refs.~\cite{alphas1,alphas2,alphas14} 
we detailed the merits and assumptions of our strategy,
and commented on the weaknesses of the truncated-OPE strategy as 
a motivation for the design of this alternate approach.

Reference~\cite{Pich} presented a detailed analysis of the determination of
$\a_s$ based on the truncated-OPE strategy. 
This allowed us to devise a series of explicit
tests to assess the validity of this approach; these are 
discussed in Secs.~\ref{PRresults} and~\ref{fake data}. 
These tests demonstrate unambiguously that the truncated-OPE strategy fails 
to produce reliable results for $\a_s(m_\t^2)$. In Sec.~\ref{PRresults}
we showed that varying the input values assumed for the higher 
dimension OPE condensates, $C_D$, in the truncated-OPE strategy 
can lead to significantly lower values for $\a_s(m_\t^2)$, about 8\% lower 
in our example. This should be compared to the $<4\%$ total error 
claimed in Ref.~\cite{Pich}. In Sec.~\ref{fake data}, we showed that the 
truncated-OPE strategy is incapable of detecting residual 
DVs,\footnote{By ``residual,'' we refer to those integrated DV 
contributions that remain even after the partial suppression produced by 
the use of pinched weights.} even when applied to a fake data set 
known to include them.  As a consequence, it is incapable of 
reproducing the correct result for $\a_s$ in such a situation, 
finding instead a value about 7\% too high in our fake-data test, 
and more than 5~$\s$ away from the true value on which the fake data 
is based. It is to be noted that 7-8\% deviations are larger 
than the differences between the CIPT and FOPT values of the coupling. 
These failures result from shortcomings in the two main 
assumptions on which the truncated-OPE strategy is based:
(1) the assumption that a number of higher-dimension OPE contributions
can safely be set to zero by hand, for which there is no basis in QCD; 
(2) the assumption that DVs can be effectively neglected, even though the 
data clearly show resonance effects which, in the context of the FESR 
analysis, necessarily produce some level of quark-hadron duality
violation. Our conclusion is that the truncated-OPE-model strategy does not
hold up to detailed scrutiny, and should no longer be used for a 
precision determination of $\a_s(m_\t^2)$ from hadronic $\t$ decays.  

Of course, the alternative DV-model strategy should be subjected to similar
scrutiny. Since this strategy was not the main topic of this article, we 
did not review the complete analysis of its application to the ALEPH data 
presented already in Ref.~\cite{alphas14}, in which many consistency tests 
were carried out (our result for $\a_s(m_\t^2)$ is based on six different 
fits, and our analysis satisfies several $V-A$ sum rules within errors). 
Moreover, the fits reported in Tables~\ref{tab11} to \ref{tab14} in 
Sec.~\ref{PRresults} are fully consistent with the results of Ref.~\cite{alphas14}, 
and thus provide further evidence for the stability of results 
obtained using the DV-model strategy. We also, in Sec.~\ref{criticism}, 
refuted the criticism of Ref.~\cite{alphas14} contained in Sec.~7 of
Ref.~\cite{Pich}, showing that the evidence on which this criticism 
was presumed to be based, in fact, actually provides further support for the 
validity of the DV-model strategy. In Sec.~\ref{DV-model}
we contrasted the two strategies, showing how much better the
DV-model strategy performs than the truncated-OPE strategy. 

We are of course
aware of the fact that the need to include a parametrization of
DVs on the theory side of the FESR analysis of spectral functions obtained
from hadronic $\t$ decays leads to a certain model dependence in the
strategy we employ. This is, however, true of {\it both} strategies: 
in the DV-model strategy, this is explicit, through the use of
Eq.~(\ref{ansatz}), while in the truncated-OPE-model strategy, it is implicit, 
through the neglect of DVs, which corresponds to the {\it ansatz}  
$\r^{\rm DV}_{V/A}(s)=0$, $s\ge s_0$, and the uncontrolled truncation 
of the OPE. We reiterate that the investigation
presented here shows unambiguously that the truncated-OPE-model fails
to yield reliable results, and presents further evidence of the 
robustness of the DV-model approach. More precise data for the $V$ 
and $A$ spectral functions, possibly obtainable from $\t$-decay data 
collected at BaBar and/or Belle, or from future 
$e^+e^-\to\mbox{hadrons}$ cross-section data, would make it 
possible to subject the DV-model strategy to more stringent tests than 
possible at present, using only OPAL and ALEPH data.

It has been pointed out that the results of Ref.~\cite{alphas14} for 
$\a_s(m_\t^2)$, found using the revised ALEPH data \cite{ALEPH13},
lie about $0.03$ below those of Ref.~\cite{alphas2}, found (using
the same strategy) from the OPAL data \cite{OPAL}. Assuming the 
two data sets to be uncorrelated, and largely because of the larger errors
on the OPAL-based result, the difference amounts to about 1.4\ $\s$.
This means that the two determinations are consistent with each other, 
and thus that it is appropriate to consider their weighted average, as was
done in Ref.~\cite{alphas14}. This yields $\a_s(m_\t^2)=0.303(9)$ (FOPT) and
$\a_s(m_\t^2)=0.319(12)$ (CIPT), results consistent with those
of a recent, preliminary combined fit of the ALEPH and OPAL data reported 
in Ref.~\cite{mainz}.

In order to compare this result with that of the truncated-OPE-model 
strategy, we have to resort to the original OPAL analysis of the OPAL 
data \cite{OPAL}, because Ref.~\cite{Pich} did not consider the OPAL data. 
In view of this, the
best one can do is to combine the results of Ref.~\cite{Pich}, 
$\a_s(m_\t^2)=0.319(12)$ (FOPT) and $\a_s(m_\t^2)=0.335(13)$ (CIPT), with 
those of Ref.~\cite{OPAL}, $\a_s(m_\t^2)=0.324(15)$ (FOPT) and
$\a_s(m_\t^2)=0.348(21)$ (CIPT). The weighted average between Ref.~\cite{Pich}
and Ref.~\cite{OPAL} is $\a_s(m_\t^2)=0.321(9)$ (FOPT) and
$\a_s(m_\t^2)=0.339(11)$ (CIPT). Again, these values, following from the
truncated-OPE-model strategy, are about 0.02 larger than those following from
the DV-model strategy, a difference comparable to that
between the CIPT and FOPT determinations.

Without including OPAL data, the results from the two strategies are
\begin{eqnarray}
\label{finalBetal}
\a_s(m_\t^2)&=&0.296(10)\qquad\mbox{(FOPT,\ DV-model strategy \cite{alphas14})}\ ,
\\
\a_s(m_\t^2)&=&0.310(14)\qquad\mbox{(CIPT,\ DV-model strategy \cite{alphas14})}\ ,
\nonumber\\
\nonumber\\
\label{finalPR}
\a_s(m_\t^2)&=&0.319(12)\qquad\mbox{(FOPT,\ truncated-OPE-model strategy \cite{Pich})}\ ,\\
\a_s(m_\t^2)&=&0.335(13)\qquad\mbox{(CIPT,\ truncated-OPE-model strategy \cite{Pich})}\ .\nonumber
\end{eqnarray}
In this case, the difference between the results of the two strategies 
is about 0.024, larger than the roughly 0.015 difference between the 
individual CIPT and FOPT results. In this article, we have demonstrated 
that the values in Eq.~(\ref{finalPR}) result from the use of a flawed 
fitting strategy, and provided further evidence for the reliability of the 
values in Eq.~(\ref{finalBetal}). The unreliable treatment of non-perturbative 
effects in the truncated-OPE-model strategy affects both the central value 
and uncertainties of $\a_s(m_\t^2)$. As such, we believe it is 
inappropriate to include values such as those in Eq.~(\ref{finalPR}), obtained 
using this strategy, in any average for $\a_s(m_\t^2)$.

\vspace{3ex}
\noindent {\bf Acknowledgments}
\vspace{3ex}

MG would like to thank Claude Bernard for discussions, the Instituto de 
F{\'\inodot}sica de S{\~a}o Carlos of the Universidade de S{\~a}o Paulo 
for hospitality and FAPESP for partial support. This material is based 
upon work supported by the U.S. Department of Energy, Office of Science, 
Office of High Energy Physics, under Award Number DE-FG03-92ER40711. 
The work of DB is supported by the S{\~a}o Paulo Research Foundation 
(FAPESP) Grant No. 2015/20689-9 and by CNPq Grant No. 305431/2015-3.
KM is supported by a grant from the Natural Sciences and Engineering Research
Council of Canada.  SP is supported by CICYTFEDER-FPA2014-55613-P, 2014-SGR-1450. 


\end{document}